# Dipolarly-Coupled Chaotic Quantum Spin Systems

A Senior Honors Thesis
Submitted to the faculty
in partial fulfillment of the requirements for the
Degree of Bachelor of Arts in Physics

by

Dhrubo Jyoti

DARTMOUTH COLLEGE
Hanover, NH
June 2011

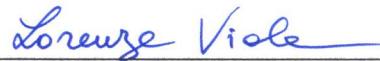
Lorenza Viola, Advisor

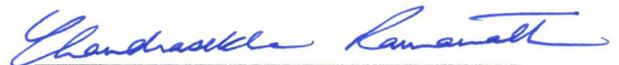
Chandrasekhar Ramanathan

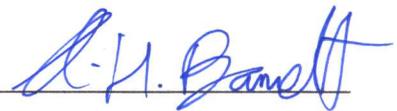
Alexander Harvey Barnett

# Abstract

We numerically explore chaoticity properties of homonuclear spin systems interacting via the so-called secular dipolar Hamiltonian. Two geometries are considered: (i) an open chain with 19 spins, (ii) a face-centered cubic lattice with 14 spins. Analysis of the energy level-spacing distribution after unfolding the energy spectrum of one symmetry sector indicates that the three-dimensional geometry is highly chaotic, while the one-dimensional system is an intermediate case for small chains but tends to higher chaoticity for larger chains. We also characterize the statistical properties of energy eigenvectors, and one- and two-body local observables. Finally, we present some preliminary results on local spin dynamics and thermalization.

# Acknowledgements

I am grateful to my advisor, **Professor Lorenza Viola**. Her vision and guidance over the course of this research, spanning my Junior and Senior years at Dartmouth, has, I feel, truly extended the dimensions of my knowledge and understanding. I am indebted to **Dr. Winton Brown** for countless hours of guidance and consultation on this project. **Professor Alex Barnett** originally inspired me to study chaos, and in particular quantum chaos. Finally, I extend my warm regards to the **Neukom Institute for Computational Science at Dartmouth** and the **Paul K. Richter and Evalyn E. Cook Richter Memorial Fund**, who jointly sponsored my leave-term, on-campus research on this project during Summer Term 2010.

*Dedicated to my younger brother Deepta, whom I love above all else, my parents, and my late music teacher, Ustad Sheikh Sattar Muhammad.*

# Table of Contents

# Introduction

Most of us are familiar with the common metaphor for chaos, the butterfly effect: tiny happenings – the flapping of a butterfly's wing in Brazil – may in time have significant consequences – altering the path of a tornado in Texas; **classical chaos** is defined as the exponential divergence of infinitesimally perturbed trajectories [1]. Ever since Poincaré's pioneering work in the late 19th century, chaos has been shown to exist and play an important role in a variety of systems and processes, ranging from simple hand-held objects (*e.g.,* a double-pendulum) to astronomical bodies (*e.g.,* the Sun-Earth-Moon system, asteroids, or even the motion of planets) to complex macroscopic phenomena (*e.g.,* from the milk mixing in our coffee to financial markets, cardiac rhythms, and so on). In particular, from a physical standpoint, the assumption of sufficiently complex dynamics (so-called "molecular chaos assumption" by Boltzmann [1]) plays a crucial role in determining the emergence of well-defined equilibrium properties in statistical approaches to many-body systems.

The predictability and tractability of regular, *i.e.,* non-chaotic processes is in striking contrast with the irregularity and apparent randomness ("pseudo-randomness") of chaotic phenomena. Imagine the familiar smooth, periodic motion of a simple pendulum; compare that with the trajectory of a chaotic system such as the double pendulum (shown in the Figure) which may be reminiscent of a toddler's first experience with pen and paper.

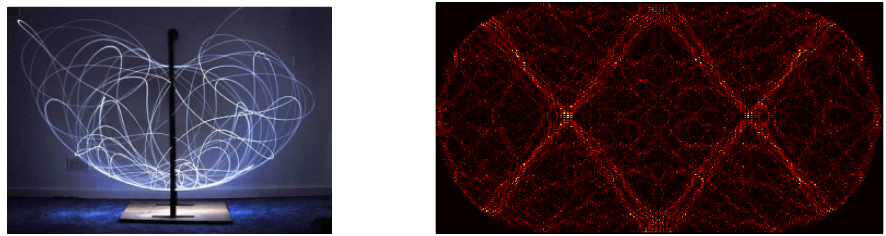

Figure 1: **(**Left**)** Long exposure photograph of the path of a double pendulum (Wikipedia) (Right) Period-4 scars in stadium billiard (from my Math 53: *Chaos!* Final Paper).

Most large-scale real-world systems show some degree of chaoticity; hence, the study of chaos is motivated not only by theoretical interest but also by practical relevance. But what about the world of the very small and invisible? How can we understand and explore chaos at the molecular and atomic levels? These length scales are the domain of quantum mechanics. How do we even think of quantum "chaoticity" as opposed to "regularity," given that the evolution of an isolated quantum system is unitary, and thus no exponential divergence of trajectories can be meaningful in the first place? Aside from the fundamental theoretical interest in these questions, it is important to acknowledge that due to the perpetual drive towards miniaturization, quantum mechanics is becoming ever more important in our everyday technologies; thus, it is imperative to develop better theoretical and experimental tools to probe chaoticity at the deepest levels of Nature. This is the subject of **quantum chaos**.

Shown above is a wave-function of the stadium billiard, a traditional problem in so-called "wave chaos". Billiards are a paradigmatic example of dynamical systems which possess a natural classical counterpart. For such systems, there is consensus among physicists that the **non-separability** of the governing dynamical equation (*i.e.,* the Hamilton-Jacobi equation in the classical limit and the Schrodinger equation in the quantum case, respectively) does provide "the deepest characterization of chaos" [2]. Not all quantum-dynamical systems admit, however, a natural classical counterpart. A chief example is provided by many-body systems of interacting quantum degrees of freedom, such as localized spins, indistinguishable fermions or bosons, and so on. A generalized definition of chaos applicable to both the macroscopic and the microscopic worlds – including those dynamical systems *without* a natural classical counterpart – is yet to be formulated with sufficient generality and full mathematical rigor.

Literature on the subject (see, for example, [2, 3, 4, 5]) has established that apparent randomness can emerge in the spectral (eigenvalue and eigenvector) properties of the quantum Hamiltonian, motivating a definition of "non-integrability" and quantum chaos based on the impossibility to efficiently "diagonalize" the relevant Hamiltonian and, consequently, the need to invoke a "coarse-grained" statistical description. In particular, such chaotic quantum systems closely obey predictions from Random Matrix Theory (RMT), a statistical approach conceived in the 1950s by E. P. Wigner and others to study complex nuclei. Specifically, the spectral statistics of Hamiltonian matrices of chaotic systems with time-reversal symmetry have been shown to resemble those of matrices belonging to the *Gaussian Orthogonal Ensemble* (GOE)[†] [2].

Since *generic* quantum many-body Hamiltonians will, typically, be chaotic, obtaining a better characterization of their statistical properties as the number of degrees of freedom (*e.g.*, the number of subsystems, $N$) grows is of both fundamental and practical importance. In particular, recent experimental progress in controlling the dynamics of isolated quantum systems has recently prompted researchers to carefully scrutinize the mechanisms by which well-defined thermodynamic values can emerge for quantum observables in the long-time limit, and to explore the role played by chaoticity in the resulting *thermalization* [6].

Our goal in this thesis will be to take some steps toward improving our understanding of these broad questions, by focusing on a simple yet rich and physically relevant many-body quantum system, namely a system of dipolarly coupled spin-1/2 degrees of freedom. The exact dipolar interaction has been previously studied in [8]; we will be focusing on an approximation of the dipolar interaction (valid in the limit of a strong external magnetic field): the so-called **secular dipolar interaction**, which we proceed to introduce and describe next.

---

[†] A GOE matrix is composed of normally-distributed random numbers with standard deviation 1 off the diagonal and $\sqrt{2}$ on the diagonal, and mean 0 for both cases.

# Chapter 1

# Secular Dipolar Hamiltonian

This thesis focuses on the dipolar interaction of several homonuclear spin-1/2 nuclear spins (corresponding to "qubits" in quantum information theory terminology) placed in a strong static uniform magnetic field. In this chapter, we introduce our model, beginning with the simple scenario of two spins interacting as dipoles and working our way up constructing the formalism required to solve this problem using quantum theory.

## 1.1 Two-Spin Dipolar Interaction

A particle with spin angular momentum $\boldsymbol{S}$ has magnetic dipole moment

$$\boldsymbol{\mu} = \gamma \boldsymbol{S}$$

where $\gamma$ is the gyromagnetic ratio of the particle. The magnetic field due to a dipole $\boldsymbol{\mu_i}$ is given by

$$\boldsymbol{B_i}(\boldsymbol{r_{ij}}) = \frac{\mu_0}{4\pi r_{ij}^3}\left[3(\boldsymbol{\mu_i} \cdot \hat{r}_{ij})\hat{r}_{ij} - \boldsymbol{\mu_i}\right]$$

If we introduce another dipole $\boldsymbol{\mu_j}$, then the interaction Hamiltonian is given by

$$H_D = -\boldsymbol{\mu_j} \cdot \boldsymbol{B_i}$$

Substituting in for $\boldsymbol{B_i}$, we obtain the expression for the interaction energy of two magnetic dipoles, the so-called "dipolar" Hamiltonian,

$$H_D = \frac{\mu_0}{4\pi r_{ij}^3}\left[\boldsymbol{\mu_i} \cdot \boldsymbol{\mu_j} - 3(\boldsymbol{\mu_i} \cdot \hat{r}_{ij})(\boldsymbol{\mu_j} \cdot \hat{r}_{ij})\right]$$

When the two dipoles are placed in an external magnetic field $\boldsymbol{B_{ext}}$, a Zeeman component $H_z$ is introduced to the full interaction Hamiltonian, which is now given by

$$H = H_Z + H_D$$

where

$$H_Z = \sum_{i=1}^{2} B_{ext}\sigma_z^i$$

Note that the expectation value of their spins $\langle \boldsymbol{S_i} \rangle$ and $\langle \boldsymbol{S_j} \rangle$ will precess around the direction of $\boldsymbol{B_{ext}}$, as shown in **Figure 2**.

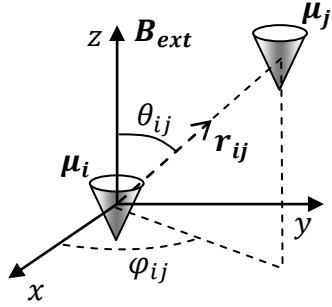

Figure 2: Two nuclear spins in an external magnetic field that defines the $\hat{\boldsymbol{z}}$ ('quantization') axis.

Let us adopt the shorthand notation $\theta = \theta_{ij}$, $\varphi = \varphi_{ij}$. Then, using $\hat{r}_{ij} = (sin\theta cos\varphi, sin\theta sin\varphi, cos\theta)$, we have

$$\boldsymbol{\mu_i} \cdot \hat{r}_{ij} = \frac{\gamma\hbar}{2}\boldsymbol{\sigma_i} \cdot \hat{r}_{ij} = \frac{\gamma\hbar}{2}\left(\sigma_i^x sin\theta cos\varphi + \sigma_i^y sin\theta sin\varphi + \sigma_i^z cos\theta\right)$$

where we have the Pauli matrices

$$\sigma_i^x = \begin{pmatrix} 0 & 1 \\ 1 & 0 \end{pmatrix}; \quad \sigma_i^y = \begin{pmatrix} 0 & i \\ -i & 0 \end{pmatrix}; \quad \sigma_i^z = \begin{pmatrix} 1 & 0 \\ 0 & -1 \end{pmatrix}$$

Hence,

$$(\boldsymbol{\mu}_i \cdot \hat{r}_{ij})(\boldsymbol{\mu}_j \cdot \hat{r}_{ij})$$

$$= \left(\frac{\gamma\hbar}{2}\right)^2 \Big[\sigma_i^x\sigma_j^x \sin^2\theta\cos^2\varphi + \sigma_i^y\sigma_j^y \sin^2\theta\sin^2\varphi + \sigma_i^z\sigma_j^z \cos^2\theta$$
$$+ \left(\sigma_i^x\sigma_j^y + \sigma_i^y\sigma_j^x\right)\sin^2\theta\sin\varphi\cos\varphi$$
$$+ \left(\sigma_i^y\sigma_j^z + \sigma_i^z\sigma_j^y\right)\sin\theta\cos\theta\sin\varphi$$
$$+ \left(\sigma_i^x\sigma_j^z + \sigma_i^z\sigma_j^x\right)\sin\theta\cos\theta\cos\varphi\Big]$$

Lastly, we replace $\sigma^x$ and $\sigma^y$ with spin-creation ($\sigma^+$) and annihilation ($\sigma^-$) operators using the identities $\sigma^x = \sigma^+ + \sigma^-$ and $\sigma^y = i(\sigma^- - \sigma^+)$. After some algebra,

$$(\boldsymbol{\mu}_i \cdot \hat{r}_{ij})(\boldsymbol{\mu}_j \cdot \hat{r}_{ij})$$

$$= \left(\frac{\gamma\hbar}{2}\right)^2 \Big[(\sigma_i^+\sigma_j^+ + \sigma_i^+\sigma_j^- + \sigma_i^-\sigma_j^+ + \sigma_i^-\sigma_j^-)\sin^2\theta$$
$$- 2(\sigma_i^+\sigma_j^+ + \sigma_i^-\sigma_j^-)\sin^2\theta\sin^2\varphi + \sigma_i^z\sigma_j^z\cos^2\theta$$
$$+ 2i(\sigma_i^-\sigma_j^- - \sigma_i^+\sigma_j^+)\sin^2\theta\sin\varphi\cos\varphi$$
$$+ i(\sigma_i^-\sigma_j^z - \sigma_i^+\sigma_j^z + \sigma_i^z\sigma_j^- - \sigma_i^z\sigma_j^+)\sin\theta\cos\theta\sin\varphi$$
$$+ (\sigma_i^+\sigma_j^z + \sigma_i^-\sigma_j^z + \sigma_i^z\sigma_j^+ + \sigma_i^z\sigma_j^-)\sin\theta\cos\theta\cos\varphi\Big]$$

Next, we substitute in $\cos\varphi = \frac{1}{2}(e^{i\varphi} + e^{-i\varphi})$ and $\sin\varphi = \frac{1}{2i}(e^{i\varphi} - e^{-i\varphi})$.

$$(\boldsymbol{\mu}_i \cdot \hat{r}_{ij})(\boldsymbol{\mu}_j \cdot \hat{r}_{ij})$$

$$= \left(\frac{\gamma\hbar}{2}\right)^2 \Big[(\sigma_i^+\sigma_j^- + \sigma_i^-\sigma_j^+)\sin^2\theta + (\sigma_i^-\sigma_j^- e^{2i\varphi}$$
$$+ \sigma_i^+\sigma_j^+ e^{-2i\varphi})\sin^2\theta + \sigma_i^z\sigma_j^z\cos^2\theta$$
$$+ (\sigma_i^-\sigma_j^z e^{i\varphi} + \sigma_i^z\sigma_j^- e^{i\varphi} + \sigma_i^+\sigma_j^z e^{-i\varphi} + \sigma_i^z\sigma_j^+ e^{-i\varphi})\sin\theta\cos\theta\Big]$$

Note also that $\boldsymbol{\mu}_i \cdot \boldsymbol{\mu}_j = \mu_i^x\mu_j^x + \mu_i^y\mu_j^y + \mu_i^z\mu_j^z = \left(\frac{\gamma\hbar}{2}\right)^2 \left[2(\sigma_i^+\sigma_j^- + \sigma_i^-\sigma_j^+) + \sigma_i^z\sigma_j^z\right]$. Substituting the expressions for $(\boldsymbol{\mu}_i \cdot \hat{r}_{ij})(\boldsymbol{\mu}_j \cdot \hat{r}_{ij})$ and $\boldsymbol{\mu}_i \cdot \boldsymbol{\mu}_j$ into $H_D$, we finally obtain the full dipolar Hamiltonian

$$H_D = \frac{\mu_0}{4\pi r_{ij}^3}\left(\frac{\gamma\hbar}{2}\right)^2 [A + B + C + D + E + F]$$

where we have introduced the following quantities:

$A = \sigma_i^z \sigma_j^z (1 - 3\cos^2\theta)$

$B = (\sigma_i^+ \sigma_j^- + \sigma_i^- \sigma_j^+)(3\cos^2\theta - 1)$

$C = 3(\sigma_i^+ \sigma_j^z + \sigma_i^z \sigma_j^+)\sin\theta\cos\theta e^{-i\varphi}$

$D = 3(\sigma_i^- \sigma_j^z + \sigma_i^z \sigma_j^-)\sin\theta\cos\theta e^{i\varphi}$

$E = -3\sigma_i^+ \sigma_j^+ \sin^2\theta e^{-2i\varphi}$

$F = -3\sigma_i^- \sigma_j^- \sin^2\theta e^{2i\varphi}$

## 1.2   Secular Approximation

Note that the state space for a single spin-1/2 degree of freedom is a dimension-2 Hilbert space $\mathcal{H}$ spanned by the basis $\{|\uparrow\rangle, |\downarrow\rangle\}$ where $|\uparrow\rangle$ and $|\downarrow\rangle$ represent the two possible spin states up- and down-spin, respectively, defined relative to the quantization axis $\hat{z}$ specified by the direction of the external magnetic field.

The Hilbert space $\boldsymbol{\mathcal{H}}$ for a 2-spin system is simply the tensor product of the Hilbert spaces of the individual spins, that is $\boldsymbol{\mathcal{H}} = \mathcal{H} \otimes \mathcal{H}$. Thus, $\boldsymbol{\mathcal{H}}$ is spanned by the tensor product of the bases of the two constituent Hilbert spaces, viz. $\boldsymbol{\mathcal{H}} = span(\{|\uparrow\rangle, |\downarrow\rangle\} \otimes \{|\uparrow\rangle, |\downarrow\rangle\}) = span(\{|\uparrow\rangle \otimes |\uparrow\rangle, |\uparrow\rangle \otimes |\downarrow\rangle, |\downarrow\rangle \otimes |\uparrow\rangle, |\downarrow\rangle \otimes |\downarrow\rangle\})$. The 'standard' basis for $\boldsymbol{\mathcal{H}}$ can be expressed as $\{|\uparrow\uparrow\rangle, |\uparrow\downarrow\rangle, |\downarrow\uparrow\rangle, |\downarrow\downarrow\rangle\}$ where we have used compact notation. In this basis, the six terms $A, B, C, D, E$ and $F$ from the previous section is conveniently represented by the matrix

$$H_D = \frac{\mu_0}{4\pi r_{ij}^3}\left(\frac{\gamma\hbar}{2}\right)^2 \begin{pmatrix} A & C & C & E \\ D & A & B & C \\ D & B & A & C \\ F & D & D & A \end{pmatrix}$$

In the limit of strong $B_{ext}$, *i.e.*,

$$B_{ext} \gg \frac{\mu_0}{4\pi r_{ij}^3}\left(\frac{\gamma\hbar}{2}\right)$$

we can make the following approximation, known as a *secular approximation*: The matrix elements outside the central block $\begin{pmatrix} A & B \\ B & A \end{pmatrix}$ and the diagonal represent transitions for which $|\Delta S^z| > 0$, where $S^z = S_i^z + S_j^z$ is the total angular momentum operator along the z-axis. Physically, these transitions are unlikely since the dipolar interaction strength is very small compared to the change in energy $\Delta E = \gamma B_{ext}\Delta S^z$. Hence, the C, D, E and F terms can be neglected, and the dipolar Hamiltonian reduces to

$$H_D = \frac{\mu_0}{4\pi r_{ij}^3}\left(\frac{\gamma\hbar}{2}\right)^2 (3\cos^2\theta - 1)(\sigma_i^+\sigma_j^- + \sigma_i^-\sigma_j^+ - \sigma_i^z\sigma_j^z)$$

Or, slightly more compactly:

$$H_D = \frac{\mu_0}{4\pi r_{ij}^3}\left(\frac{\gamma\hbar}{2}\right)^2 \frac{(3\cos^2\theta - 1)}{2}(\boldsymbol{\sigma_i} \cdot \boldsymbol{\sigma_j} - 3\sigma_i^z\sigma_j^z)$$

## 1.3 Generalization to $N$ qubits

Generalizing the 2-spin system to N spins is straightforward. The Hamiltonian governing the $N$-spin in a static magnetic field is, as before, the sum of a Zeeman and a dipolar component

$$H = H_Z + H_D$$

where the Zeeman component is given by

$$H_Z = \sum_i^N B_{ext}\sigma_z^i = B_{ext}\sum_i^N \sigma_z^i$$

since the magnetic field is assumed to be uniform and the spins are assumed to have the same Larmor frequency. (Under such conditions, $H_Z$ can be effectively ignored, as will be explained near the end of Section 1.6). The N-particle dipolar interaction is simply the sum of the N-choose-2 *i.e.* $_2^NC = \frac{N(N-1)}{2}$ two-body dipolar interactions. That is,

$$H_D = \sum_{i<j}^{N} d \left[ \frac{\boldsymbol{\sigma}^i \cdot \boldsymbol{\sigma}^j}{r_{ij}^3} - \frac{3(\boldsymbol{\sigma}^i \cdot \boldsymbol{r}_{ij})(\boldsymbol{\sigma}^j \cdot \boldsymbol{r}_{ij})}{r_{ij}^5} \right]$$

where

$$d = \frac{\mu_0}{4\pi} \left( \frac{\gamma \hbar}{2} \right)^2$$

As shown in the previous section, in the limit of a sufficiently strong external magnetic field, the above dipolar interaction reduces to the secular form

$$H_D \approx \sum_{i<j}^{N} c_{ij} (\boldsymbol{\sigma}^i \cdot \boldsymbol{\sigma}^j - 3\sigma_z^i \sigma_z^j)$$

where

$$c_{ij} = \frac{d}{r_{ij}^3} \left( \frac{3\cos^2 \theta_{ij} - 1}{2} \right)$$

Here, we are primarily interested in the statistical properties of energy spectra and observables; thus, it suffices to set $d = 1$. Insight is gained by replacing the Pauli spin matrices with spin-creation ($\sigma_+$) and annihilation ($\sigma_-$) operators. Once again using the identities $\sigma_x = \sigma_+ + \sigma_-$ and $\sigma_y = i(\sigma_- - \sigma_+)$, we have

$$\boldsymbol{\sigma}^i \cdot \boldsymbol{\sigma}^j - 3\sigma_z^i \sigma_z^j = \sigma_x^i \sigma_x^j + \sigma_y^i \sigma_y^j - 2\sigma_z^i \sigma_z^j = 2(\sigma_+^i \sigma_-^j + \sigma_-^i \sigma_+^j - \sigma_z^i \sigma_z^j)$$

Thus, the dipolar interaction can be finally expressed as,

$$\boxed{H_D = \sum_{i<j}^{N} c_{ij} h_{ij}} \qquad (1.1)$$

where,

$$h_{ij} = \sigma_x^i \sigma_x^j + \sigma_y^i \sigma_y^j - 2\sigma_z^i \sigma_z^j \equiv 2(\sigma_+^i \sigma_-^j + \sigma_-^i \sigma_+^j - \sigma_z^i \sigma_z^j)$$

Note that the secular dipolar Hamiltonian $H_D$ is a special case of the most general spin-1/2 Hamiltonian (the so-called Heisenberg XYZ model)

$$H_{XYZ} = -\frac{1}{2} \sum_{i<j}^{N} [J_x^{ij} \sigma_x^i \sigma_x^j + J_y^{ij} \sigma_y^i \sigma_y^j + J_z^{ij} \sigma_z^i \sigma_z^j] \qquad (1.2)$$

In particular, the secular dipolar Hamiltonian corresponds to $J_x^{ij} = J_y^{ij} \neq J_z^{ij}$, sometimes referred to as the XXZ model. The case where the Ising term vanishes ($J_z^{ij} = 0$) is, likewise, known in the condensed-matter literature as the XY model, with $J_x^{ij} = J_y^{ij}$ corresponding to the isotropic limit. Remarkably, the (secular) dipolar Hamiltonian is found naturally in many solid-state systems, such as fluorapatite (see Figure 3).

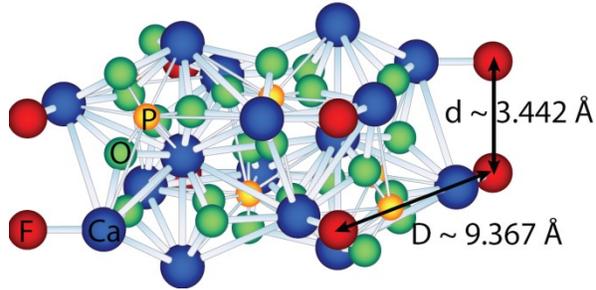

Figure 3: Schematic of the crystal *fluorapatite*; the linear arrangement of fluorine atoms has been used as a physical realization of a one dimensional spin chain for the isotropic XY Hamiltonian, obtained from the naturally occurring dipolar Hamiltonian by using suitable control processes, by Professor Ramanathan and coworkers in [8].

## 1.4 Hilbert Space

The state space for an N-site spin-1/2 system is the tensor product of the N individual state spaces, a straightforward generalization the 2-site case. In what follows, we adopt the following notation:

$$|0\rangle \equiv |\uparrow\rangle \; ; \quad |1\rangle \equiv |\downarrow\rangle$$

Accordingly, in compact form,

$$|1\rangle \otimes |0\rangle \otimes |1\rangle \ldots \otimes |1\rangle \equiv |101\ldots 1\rangle$$

Each basis state of the system may thus be represented by a string of binary digits, i.e. an N-bit long bit-string. The Hilbert space for the N-qubit system is of dimension $2^N$. A natural choice for the basis of the system is just the binary representations of the integers $\{0,1,2,3,\ldots,2^N\}$. We shall refer to this set as the **computational basis**. For example, for N=4, the computational basis is,

$$\mathcal{H} = \{ |0\rangle \equiv |0000\rangle, |1\rangle \equiv |0001\rangle, |2\rangle \equiv |0010\rangle, |3\rangle \equiv |0011\rangle, \ldots, |14\rangle \equiv |1110\rangle, |15\rangle \equiv |1111\rangle \}$$

To construct the Hamiltonian for the system, we calculate the two-body contribution $h_{ij}$ from each pair of qubits $(i,j)$ for all $i > j$. Each element of the matrix $h_{ij}$ is given by the inner product of $(\sigma_+^i \sigma_-^j + \sigma_-^i \sigma_+^j) - \sigma_z^i \sigma_z^j$ with the bra and ket of computational basis state, such as $\langle 101\ldots 0|$ and $|011\ldots 0\rangle$, respectively. The operator $\sigma_+^i \sigma_-^j + \sigma_-^i \sigma_+^j$ (the so-called "hopping term") is non-zero only when all three following conditions are met: (i) the $i^{\text{th}}$ bit of the bra is the same as the $j^{\text{th}}$ bit of the ket, as well as vice versa, (ii) the $i^{\text{th}}$ and $j$th bit of bra and ket are different, and finally (iii) all bits other than $i$ and $j$ are identical in the bra and ket. This implies that if the Hamming distance[1] between the bra and ket bit strings is not equal to 2, the hopping term is zero.

---

[1] defined as the number of bits that are different in the two bit strings. In other words, the number of 0's when we take the bitwise AND of the two bit strings

Furthermore, the Ising term is zero whenever the Hamming distance is non-zero, *i.e.* the Ising term is zero everywhere except on the diagonal. The upshot is that each two-body matrix $h_{ij}$ is sparse, and therefore the overall dipolar Hamiltonian **$H_D$ is sparse**.

## 1.5 Geometries

### I. Open Chain

The configuration we are primarily interested in is a one-dimensional open-boundary chain with $N$ equally-spaced qubits placed in a strong uniform magnetic field **$B_{ext}$** oriented perpendicular to the chain. Thus, $\theta_{ij} = \frac{\pi}{2}$ for all pairs of qubits $(i, j)$. This geometry is applicable to quasi-1D systems such as $^{19}F$ atoms in the crystal fluorapatite, as studied experimentally in [8] (shown in Figure 3 is a schematic from that paper, courtesy of Prof. Viola). Hence, the general equation for $N$-qubit secular dipolar interaction, equation (1.1), reduces to the following special case (which we will refer to as our **1D model**):

$$H_{1D} = -\frac{1}{2} \sum_{i<j}^{N} \frac{1}{r_{ij}^3} (\sigma_x^i \sigma_x^j + \sigma_y^i \sigma_y^j - 2\sigma_z^i \sigma_z^j) \qquad (1.3)$$

- **Nearest-Neighbor (NN) Approximation**

A special case of the open chain is when we consider only the so-called "nearest-neighbor couplings"; there are $N - 1$ of them, all with $r_{ij} = 1$. The Hamiltonian above then simplifies to what we will refer to as our **1D NN model**:

$$H_{1D\ NN} = -\frac{1}{2} \sum_{i}^{N-1} \left[ \sigma_x^i \sigma_x^{i+1} + \sigma_y^i \sigma_y^{i+1} - 2\sigma_z^i \sigma_z^{i+1} \right]$$

The 1D NN case is known to be *integrable* (*i.e.* it has an exact analytical solution) via an algebraic method known as the Bethe Ansatz [9].

### II. Face-Centered Cubic Lattice

The second geometry we investigate is a three-dimensional face-centered cubic lattice with $N = 14$ qubits. Just as with the open chain, the lattice is placed in a strong uniform magnetic field $\boldsymbol{B_{ext}}$ pointing in the $\hat{z}$ direction. An example of a crystal with this lattice structure is CaF$_2$.

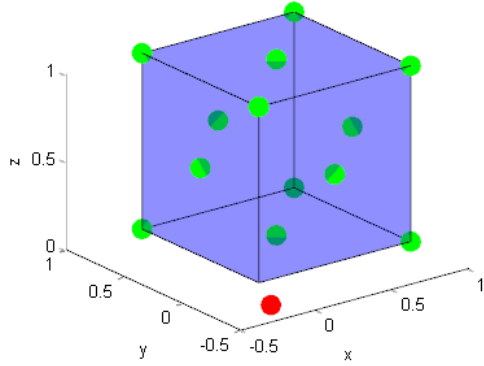

Figure 4: Face-Centered Cubic Lattice (3D model); the coordinates of red "rogue" qubit are (-0.15, -0.3, 0).

## 1.6 Block Diagonalization

$H_D$ calculated in the manner described above is a $2^N \times 2^N$ sparse matrix. Computing the eigenvalues and eigenfunctions of this full Hamiltonian would quickly become impractical for $N \approx 16$. Instead, we would like to examine just one fully-desymmetrized subspace small enough for computing eigenvalues and eigenvectors numerically within a reasonable amount of time yet large enough for a statistical description to be still meaningful. This can be achieved by reordering the basis elements in the computational basis so that $H_D$ becomes block-diagonal, with each block corresponding to a particular eigenvalue of the total spin operator, $S_z := \sum_i \sigma_z^i$. One way of doing this is to order the basis elements in ascending order of the number of 1's in their bit string. For instance, the computational basis for N=4 given above would be reordered as follows:

$$\mathcal{H} = \{\,|0\rangle \equiv |0000\rangle, |1\rangle \equiv |0001\rangle,$$
$$|2\rangle \equiv |0010\rangle, |4\rangle \equiv |0100\rangle, |8\rangle \equiv |1000\rangle, |3\rangle \equiv |0011\rangle, |5\rangle \equiv |0101\rangle,$$
$$|6\rangle \equiv |0110\rangle, |9\rangle \equiv |1001\rangle, |10\rangle \equiv |1010\rangle, |12\rangle \equiv |1100\rangle, |7\rangle \equiv |0111\rangle,$$
$$|11\rangle \equiv |1011\rangle, |13\rangle \equiv |1101\rangle, |14\rangle \equiv |1110\rangle, |15\rangle \equiv |1111\rangle\,\}$$

Next, we select one of the central blocks from this block-diagonal Hamiltonian matrix. The dimension of the subspace is given by ${}^N_kC = \frac{N!}{k!(N-k)!}$, where $k$ is the number of up or down spins in the sub-block. (Note that we are counting spins in units of 1 rather than $\frac{\hbar}{2}$; thus, for example, the subspace $S_z = +2$ refers to the block whose basis states have two more up-spins than down-spins, corresponding, in the case $N = 4$, to $\{|1\rangle \equiv |0001\rangle,\ |2\rangle \equiv |0010\rangle, |4\rangle \equiv |0100\rangle, |8\rangle \equiv |1000\rangle\}$).

Finally, note that for a given block $H^{S_z}$ of the full Hamiltonian $H = H_D + H_Z$, the Zeeman term $H_Z = B_{ext} S_z$ only adds a constant to every element on the diagonal. Formally,

$$\det(H^{S_z} - \lambda I) = \det\left(H_D^{S_z} + B_{ext} S_z I - \lambda I\right) = \det\left(H_D - \bar{\lambda} I\right) = 0$$

That is, the energy spectrum of the full Hamiltonian, $\lambda = \bar{\lambda} + S_z B_{ext}$, is simply the energy spectrum of $H_D^{S_z}$ shifted by a constant Zeeman contribution $S_z B_{ext}$. Therefore, $H_Z$ can have no effect on the behavior of our system and we shall **ignore** $H_Z$ henceforth.

## 1.7 Symmetries and Desymmetrization

Symmetries in the Hamiltonian will typically lead to degeneracies in the energy spectrum. As will be explained in the next chapter, in order to detect and measure chaoticity effectively we must identify and extract a subspace of our Hamiltonian that is not symmetric under any transformation. Such a subspace is sometimes called a *symmetry sector*.

**The 1D model is symmetric under spatial reflection.** To obtain a symmetry sector, we begin by identifying pairs of basis elements $b$ and $\tilde{b}$ that are images of each other under reflection about the center of the chain. Two lists of basic elements $\alpha$ and $\beta$ are made such that each list contains only one of $b$ and $\tilde{b}$. Note that a small minority of bit strings are invariant under reflection; these are elements are added to the list $\alpha$. Next, the two lists are concatenated and the basis elements are reordered according to the order of the combined list. This reordering constitutes a unitary transformation $U$ which is applied as follows: $\overline{H_D} = U^T H_D U$, where $\overline{H_D}$ denotes the dipolar Hamiltonian matrix in the new basis.

Finally, for each $b$ and its partner $\tilde{b}$ in the reordered basis, $b$ is mapped to $\frac{1}{\sqrt{2}}(b + \tilde{b})$ while $\tilde{b}$ is mapped $\frac{1}{\sqrt{2}}(b - \tilde{b})$, and the reflection-invariant bit strings in list $\alpha$ are mapped to themselves. This constitutes a further unitary transformation; as a result, we end up with two "desymmetrized" sub-blocks – one symmetric under reflection and the other antisymmetric. Either sub-block is now suitable to study statistical properties of eigenvalues.

**The 3D model has numerous spatial symmetries**. In order to avoid the tedious process of dealing with each one them one at a time, a simple **desymmetrization "by hand"** procedure may be used. The idea is to displace one of the qubits strategically such that all of the spatial symmetries of the system are broken, as shown in Figure 4. In particular, the results presented in this paper correspond to the position of the "rogue qubit" shown in red in the figure. In addition, the central $S_z = 0$ subspace is invariant under the $0 \leftrightarrow 1$ transformation; to avoid having to deal with this symmetry, we picked the next-central $S_z = +2$ subspace for diagonalization.

The coefficients $c_{ij}$ in the boxed expression for $H_D$ in equation (2.2) incorporate the lattice geometry. Note the sharp inverse-cubic dependence on qubit separation; a crucial implication is that when setting up the qubit geometry, one must avoid creating "clusters" of qubits. Otherwise, each cluster would essentially form a separate (nearly) isolated sub-system; energy eigenvalues of the entire system would then approximately be the sum of the energies of each cluster. Just as with symmetries, this has the undesirable effect of obscuring the statistical features of the energy spectrum that characterize chaoticity, as will be explained in detail the next chapter. The entire block diagonalization and desymmetrization procedure is illustrated pictorially in Figure 5.

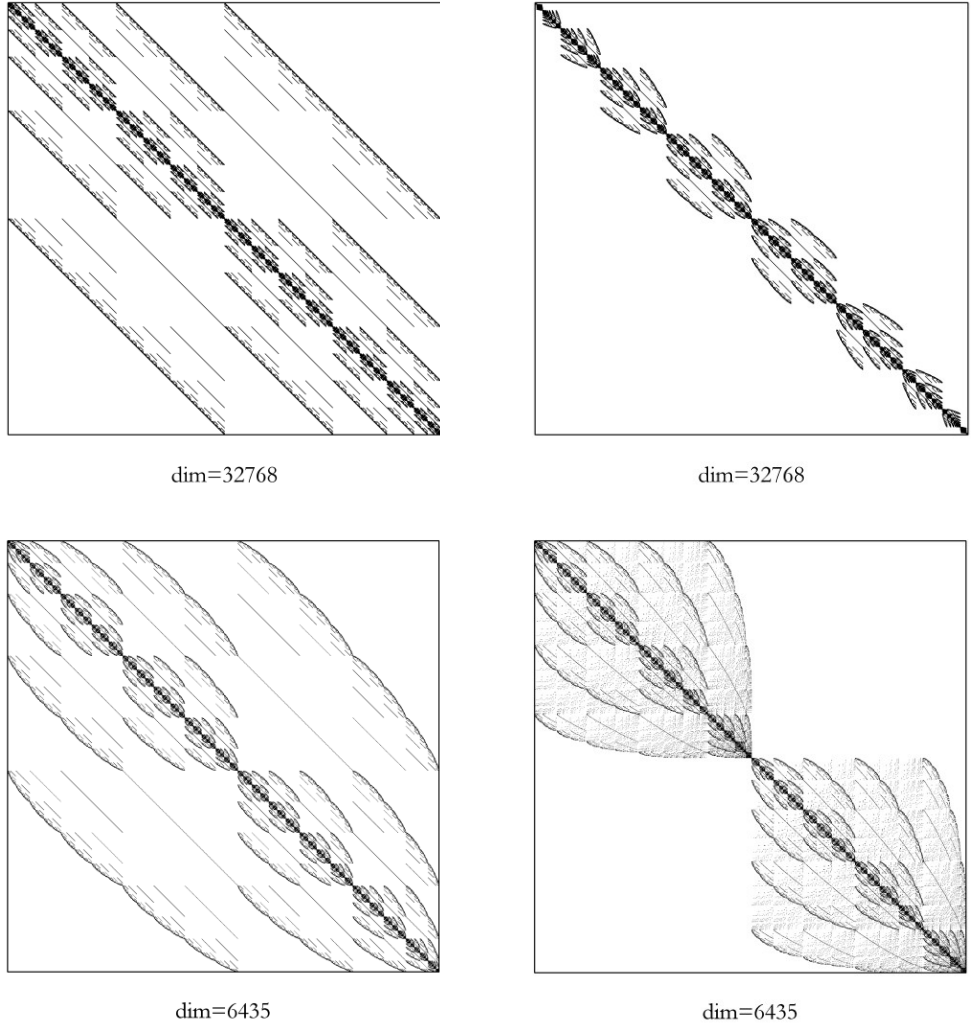

Figure 5: Sparsity plots created using the Matlab routine *spy* for the 1D model for N=15: (i) The full Hamiltonian $H_D$, $dimension = 2^{15}$, in the computational basis. (ii) $H_D$ in a reordered basis in which it is block diagonal. (iii) Zooming in on $S_z = +1$ block. (iv) The $S_z = +1$ block desymmetrized into two sub-blocks; the top sub-block is symmetric under reflection while the bottom is antisymmetric. Either one is a symmetry sector hence suitable for eigenvalue statistics (we will be using the symmetric one in the next chapter).

# Chapter 2

# Spectral Properties

## 2.1 Level Density

We begin our study of spectral properties of the secular dipolar interaction with an inspection of energy level density of a single symmetry sector. In the thermodynamic limit (*i.e.* $N \to \infty$), it is possible to show that the level density $\Omega(\tilde{E})$, where $\tilde{E} = E/E_\sigma$ is the energy spectrum normalized to have standard deviation one, should approach the normal distribution [5]. Indeed, we found the level density to be close to a Gaussian shape even for a relatively small chain of spins, as shown in Figure 6.

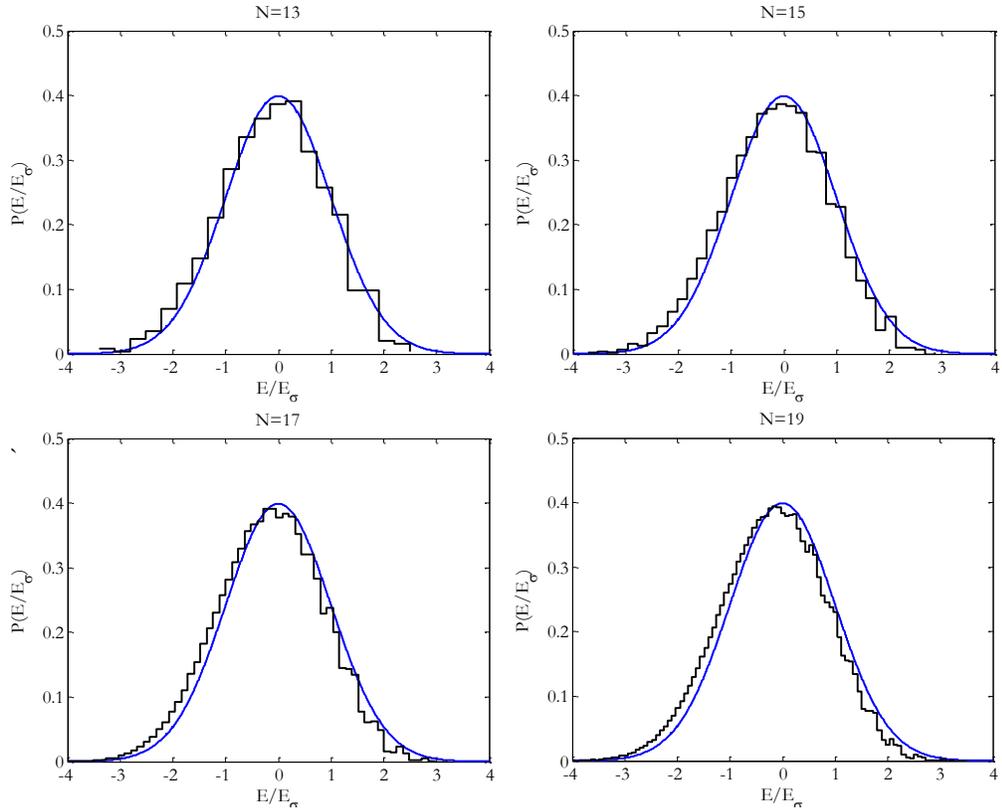

Figure 6: Histograms showing level density of one symmetry sector of the central ($S_z = +1$) subspace of 1D models corresponding to $N = 13, 15, 17$ and $19$ qubits respectively. Each spectrum has been normalized to have standard deviation one to allow comparison with the each other and with the standard normal distribution shown in blue. The dimension of the symmetry sector is 868, 3235, 12190 and 46252, respectively.

Despite the overall adherence to the Gaussian shape, note the slight off-centeredness (towards negative energy axis) of the level density. Also note the marked jaggedness for $E > 0$, as opposed to the relative smoothness for $E < 0$. This apparent asymmetry in behavior about the center of the spectrum for the 1D model shows up in several of our results as will be discussed in the course of the thesis.

## 2.2 Unfolding

A well-established spectral signature of chaos is that the energy levels of a chaotic Hamiltonian seem to "repel" each other to avoid a crossing. [2] Such a "level repulsion" results in nearby energy levels being correlated *locally*, *i.e.* the spacings between adjacent energy levels within a small enough energy window follow a characteristic distribution. The particular distribution depends on the symmetry properties of the Hamiltonain; a time reversal-invariant Hamiltonian such as our secular dipolar interaction is expected to obey (approximately) the so-called Wigner-Dyson (WD) distribution, given by

$$P_{WD}(s) = \frac{\pi}{2} s e^{-\frac{\pi}{4}s^2}$$

which is the level-spacing distribution obeyed by eigenvalue-spacings of the so-called Gaussian Orthogonal Ensemble (GOE) matrices, according to Random Matrix Theory (RMT) [2]. In contrast, spacings of integrable systems are uncorrelated; if the integrable system is sufficiently complicated, the level spacings follow the Poisson distribution [2]:

$$P_P(s) = e^{-s}$$

Note that since the density of energy levels varies across the spectrum (peaking at the center and diminishing away from the center), the *average* spacing between adjacent energy levels varies as a function of energy. To compute the level-spacing distribution for the entire spectrum, each spacing has to be normalized by dividing it with the average spacing at that location in the spectrum. This is the so-called "**unfolding**" procedure.

Although the histograms given in Figure 6 do provide a rough approximation to the level densities of the respective systems, directly using them to unfold the spectra is inadvisable since their sharp edges would distort the spacing statistics. Finer binning will not circumvent this problem since the level density as a function of energy derived from a finite, discrete spectrum is inherently noisy. In order to compute the spacing distribution accurately, what we require is the smoothly-varying component of this noisy function.

Conveniently, I found the Matlab kernel-smoothing routine *ksdensity* that is highly optimized for distributions that are approximately normal distribution, as is the case for our level densities. Shown in red in **Figure 7** below is the smoothed level density distribution obtained for the 3D model. Note the excellent lining-up and agreement with histogram plot, even near the tails. A simple self-written kernel smoother (see my Matlab routine *ismooth* in Appendix) that I tried initially did not perform nearly as well, becoming inaccurate near the tails. *ksdensity* must be using a somewhat sophisticated algorithm to be performing so well even at the tails of the spectrum, but since its code is proprietary we were unable to determine exactly how it works.

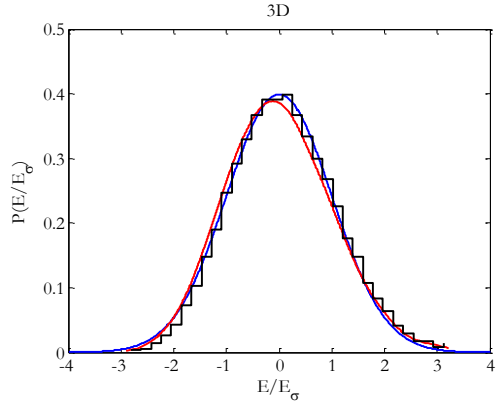

Figure 7: The level density for $S_z = +2$ block of the 3D model (black), the normal distribution (blue) and the kernel-smoothed level density produced by Matlab routine *ksdensity*.

We use this smoothed level density to unfold the spectrum as follows. If the spacing between two adjacent energy eigenvalues $E_i$ and $E_{i+1}$ is $s_i \coloneqq E_{i+1} - E_i$, the normalized spacing is given by $s_i' = \frac{s_i}{s_{avg}(E)}$ where the average spacing $s_{avg}(E)$ is equal to the reciprocal of the density in the interval $(E_{i+1}, E_i)$ which is approximately the average of the density at the end points $\frac{\Omega(E_i) + \Omega(E_{i+1})}{2}$. In summary,

$$s_i' = \frac{(E_{i+1} - E_i)\bigl(\Omega(E_i) + \Omega(E_{i+1})\bigr)}{2}$$

Figure 8 shows the spacing statistics we obtained for 1D, 3D and 1D NN models respectively after unfolding the respective spectra. The histogram for the 3D model aligns closely with the WD distribution (red), demonstrating that it clearly lies in the chaotic regime. The histogram for the 1D NN model confirms our expectation that it is integrable by closely matching up with the Poisson distribution (blue). The 1D long-range model, the one we are most interested in, appears to be in a somewhat intermediate region for small N, but for larger N it clearly approaches the fully chaotic limit.

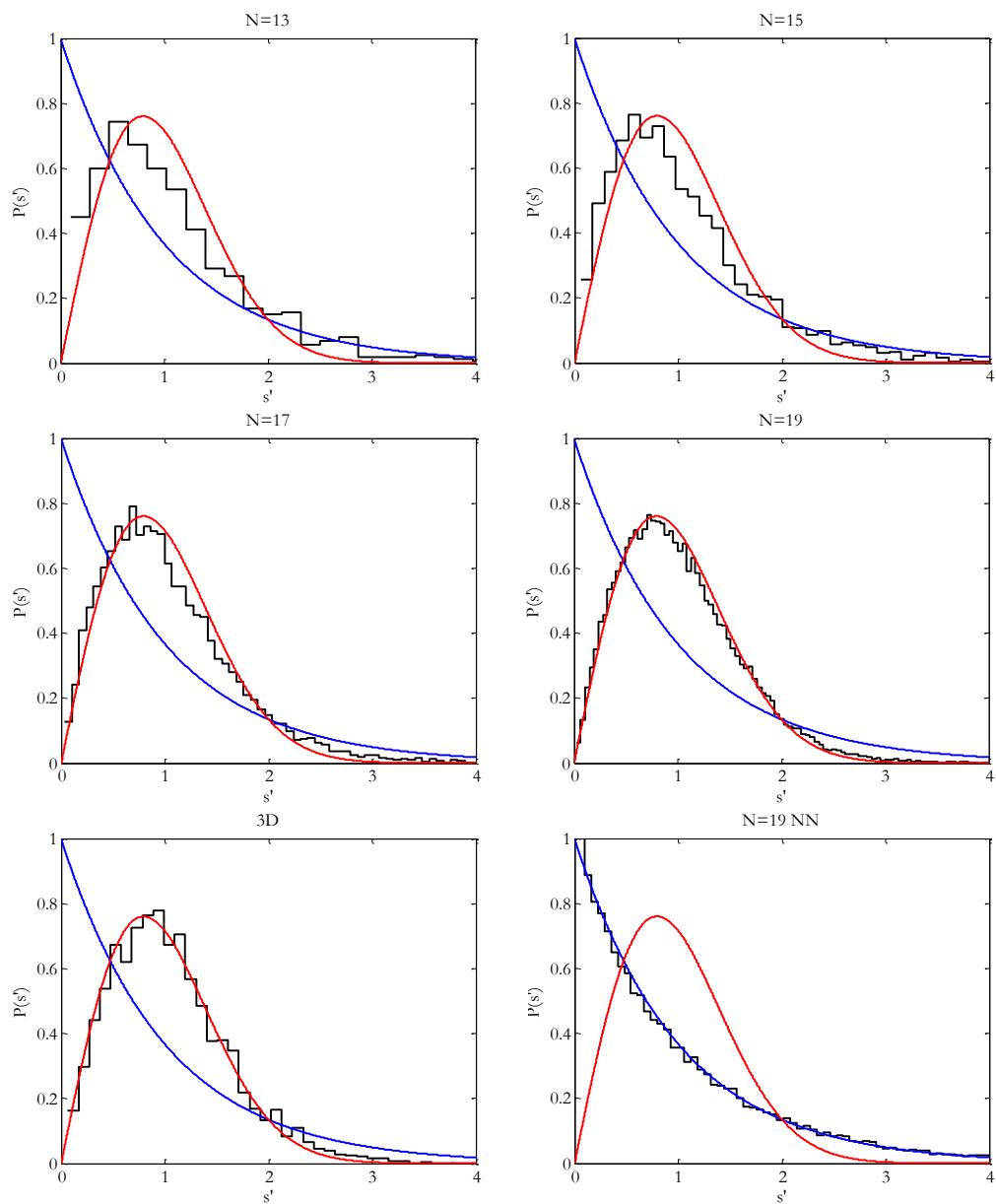

Figure 8: Level-spacing distributions of our various models.

## 2.3 Level Statistics Indicator (LSI)

A convenient way to quantify chaoticity from energy level statistics, as used for instance in [10], is the Level Statistics Indicator (LSI) defined as

$$\eta = \frac{\int_0^{s_0}[P(\tilde{s}) - P_{WD}(\tilde{s})]d\tilde{s}}{\int_0^{s_0}[P_P(\tilde{s}) - P_{WD}(\tilde{s})]d\tilde{s}}$$

where the constant $s_0 \approx 0.4729$ is the first point of intersection of the WD and Poisson curves. Note that the spacing density distribution $P(\tilde{s})$ for a finite sample of $m$ normalized spacings $\tilde{s}_i$ is simply a sum of Dirac-delta spikes, *viz.*

$$P(\tilde{s}) = \frac{1}{m}\sum_{i=1}^{m} \delta(\tilde{s} - \tilde{s}_i)$$

Thus, the integral over $P(\tilde{s})$ in the expression for $\eta$ is equal to the fraction of level spacings smaller than or equal to $s_0$, while the other integrals are constants that can be readily computed analytically. The LSI for the 1D model for N=13, 15, 17 and 19 qubits was computed to be **0.52, 0.36, 0.20,** and **0.11** respectively, indicating that for smaller chains the 1D model appears to be somewhere in some intermediate regime between chaotic and integrable, but approaches high chaoticity for larger chains. The LSI for the 3D model was computed to be **0.087** implying high chaoticity. Finally, LSI for the 1D NN model with N=19 qubits was calculated to be 1.10, confirming our knowledge that it is integrable.

## 2.4 LSI as a Function of Energy

Having defined the LSI above, we can now use it as a tool to further explore the asymmetry in behavior about $E = 0$ that we first saw in the level densities of 1D models in section 2.1. The idea is that we can estimate the LSI locally at a given point $E$ in the energy spectrum by appropriately unfolding the energy levels within a small window around $E$. In other words, if we take a window of radius $\epsilon$ centered at $E$, we may define

$$\eta(E) \approx \frac{\int_0^{s_0}[P_\epsilon^E(\tilde{s}) - P_{WD}(\tilde{s})]d\tilde{s}}{\int_0^{s_0}[P_P(\tilde{s}) - P_{WD}(\tilde{s})]d\tilde{s}}$$

where the distribution of $m_\epsilon$ normalized spacings of the $(m_\epsilon + 1)$ eigenvalues in the window is given by

$$P_\epsilon^E(\tilde{s}) = \frac{1}{m_\epsilon}\sum_{j=1}^{m_\epsilon}\delta(\tilde{s} - \tilde{s}_j)$$

After spending some time experimenting with the window parameter $\epsilon$, I computed $\eta$ at each energy eigenvalue and smoothed it with a self-written simple Gaussian smoother. The plots for the 1D and 3D models are given in **Figure 9** below. We will have more to say about the smoother in the next sections; for now, it is important to note that the smoother is known to be less reliable at the ends of the spectrum due to the scarcity of data points there.

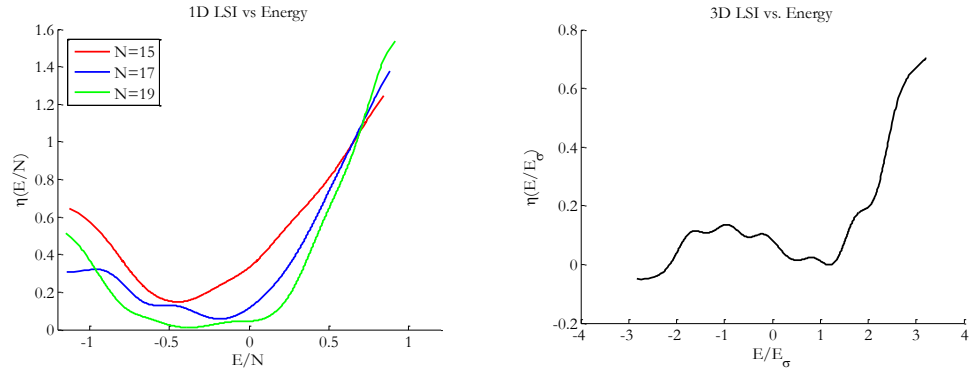

Figure 9: LSI as a function of energy for the 1D N=15, 17 and 19 models (left) and the 3D model (right).

Our theoretical expectation was that, for both models, the high and low-energy limits of the spectrum ought to be integrable so the LSI ought to be high at the opposite extremes of the energy spectrum. Figure 7 shows that the LSI does rise significantly in the high-energy end of the spectrum but not quite so much in the low-energy end. Furthermore, we expected the center of the spectrum to be the most chaotic hence lowest in LSI.

Our conjecture is that the rise in LSI in the negative extreme of the spectrum is too sudden and sharp for our relatively unsophisticated smoothing routine to fully capture. However, the fact that the most chaotic region is **not** exactly at the center of the spectrum is more difficult to explain. It suggests that the chaoticity properties of the dipolar models may have more texture to them than we may have anticipated, as we will see in chapters 3 and 4.

## 2.5 Eigenvectors

Next, we'd like to study the statistical properties of eigenvectors. Specifically, we would like to investigate how close the eigenvector components are to being maximally random (*i.e.* following normal distribution).

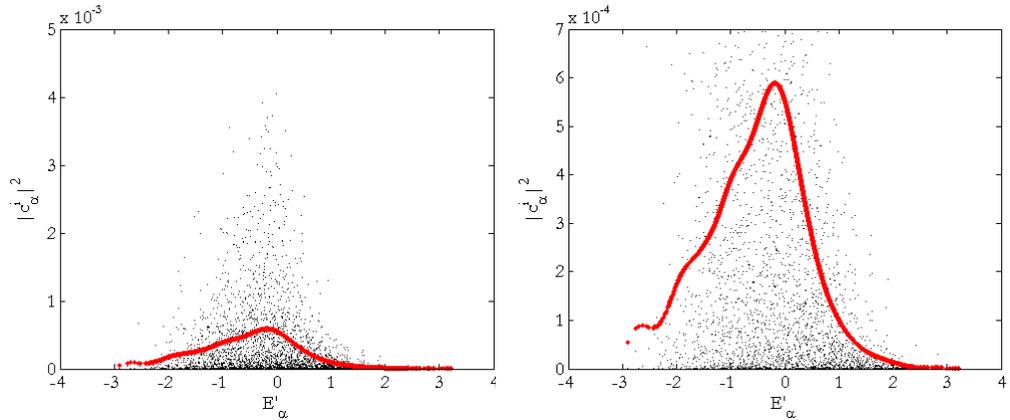

Figure 10: (Left) Component-square versus Energy (normalized to have standard deviation 1) for the 3D model for a given basis element $|i = 1925\rangle$. (Right) Diagram on the left magnified.

The black dots in Figure 10 above show the component-squares $|c_i^\alpha|^2$ plotted against energy eigenvalues $E^\alpha$ for the 3D model, where $c_i^\alpha = \langle i | E^\alpha \rangle$ for a given basis state $|i\rangle$. However, note that the variance $v_i$ of $c_i^\alpha$ for a given $i$ is known to be an arbitrary function of energy, *i.e.* $v_i(E^\alpha) = f(E^\alpha)$. Before we can find the distribution obeyed by these $c_i^\alpha$ we must normalize them – similar to the unfolding procedure that we did for eigenvalue-spacing statistics – so that $v_i$ is equal to one for all $c_i^\alpha$.

Note however that, unlike the unfolding procedure for eigenvalues where the level density was always close to a Gaussian, the functional dependence of $v_i$ with respect to energy is some arbitrary function. Matlab's *ksdensity*, which is optimized for normal distribution, may no longer be the best option. Although *ksdensity* is convenient in that it does not require a window parameter input, and that its more sophisticated algorithm usually allows a somewhat more accurate determination of the smoothly-varying trend, the routine is known to fail (and we have observed it failing upon experimentation) in cases when the data that needs to be smoothed does not decay to zero on either extreme of the x-axis.

We therefore implemented our own kernel smoothing routine that accepts a window parameter and employs a simple Gaussian kernel. An obvious advantage of our custom-written kernel smoother is that we know exactly how it works, and in our experience it has worked adequately well for most of our needs. The drawbacks are that it tends to get somewhat inaccurate near the tails, and that we have to feed it with a suitable empirically-determined window parameter. The red curve in Figure 10 shows the smoothly-varying $f(E^\alpha)$ extracted from the black dots representing $|c_i^\alpha|^2$. We then find the normalized components $\widetilde{c_i^\alpha} = \frac{c_i^\alpha}{\sqrt{f(E'_\alpha)}}$ and sample them in a histogram, as shown in Figure 11.

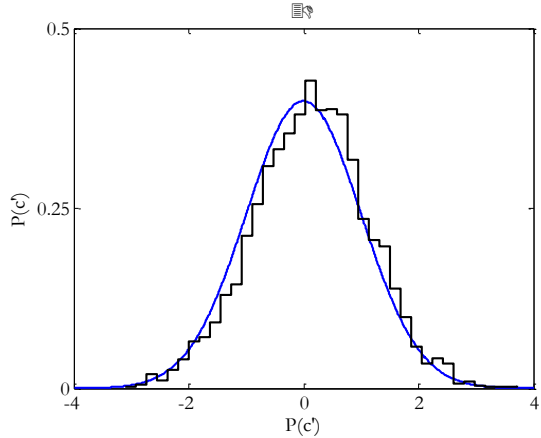

Figure 11: Normalized component statistics for the 3D model corresponding to the basis state in Figure 10.

We would like to repeat this normalization procedure for multiple basis elements and sample all of them collectively. The greater the number of components we sample, the more globally-representative and comprehensive our component distribution will be. This is our next goal, but first we must introduce a necessary tool.

## 2.6 Number of Principal Components (NPC)

A well-known quantity associated with state vectors is: the number of principal components (NPC). The NPC of a pure state $|\psi\rangle$, as used in [11], is:

$$\xi(|\psi\rangle) = \left(\sum_n |\langle n|\psi\rangle|^4\right)^{-1}$$

where the set $\{|n\rangle\}$ constitutes an orthonormal basis in the Hilbert space. Eigenvectors of GOE matrices are expected [11] to have an average NPC of

$$\overline{NPC} = \frac{D+2}{3}$$

where $D$ is the dimension of the relevant Hilbert space.

Note that if $M$ is the *modal matrix* (*i.e.* matrix whose columns are the energy eigenvectors) then $M_{i\alpha} \equiv c_i^\alpha$. For each row $i$, we define the row-wise $NPC$, $NPC_i := (\sum_\alpha |c_i^\alpha|^4)^{-1}$ and similarly for each column $\alpha$ we define the column-wise NPC, $NPC_\alpha := (\sum_i |c_i^\alpha|^4)^{-1}$. Figure 12 shows $NPC_\alpha$ for the 3D and the 1D N=15 models.

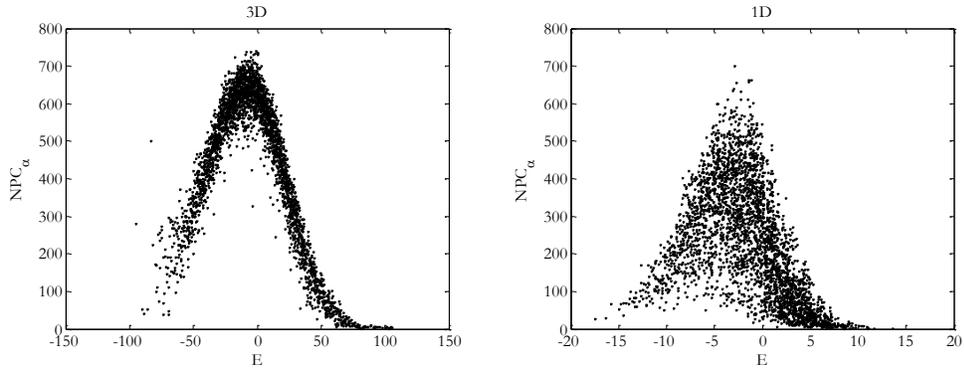

Figure 12: $\boldsymbol{NPC_\alpha}$ for the 3D (symmetry sector dimension=3003) and the 1D N=15 (symmetry sector dimension=3235) models.

Note again the asymmetry in NPC about $E = 0$, especially for the 1D model. Also note the relatively wide spread in NPC values for the 1D model. These are further hints that the chaoticity of the dipolar interaction, in particular the 1D model, has more texture than we may have originally thought.

## 2.7 Component Statistics

While experimenting with sampling the normalized components from several basis states, I found that the component distribution depended significantly upon which basis states I sampled. Specifically, the shape of the distribution depended on the range of $NPC_i$ and $NPC_\alpha$ sampled, and the dependence is significantly stronger for the 1D model than the 3D model (the 3D model was in fact relatively robust).

It was therefore difficult to settle on a set of criteria for component selection since there is not a great deal of literature on eigenvector statistics and there is no unambiguous established benchmark on component sampling. Instead, I proposed a set of criteria myself, based on a resource by my advisor Prof. Viola and her (then) graduate student W. G. Brown. They had determined in one of their papers on a related spin system [11] that there is an upper limit for the NPC beyond which it is impossible for the distribution to be Gaussian, given by the formula for $\overline{NPC}$ in the previous section. In that paper, they go on to sample eigenvectors with $NPC_\alpha$ within a narrow window around $\overline{NPC}$.

Following their work, I sampled **100 eigenvectors whose $NPC_\alpha$ are closest to** $\overline{NPC}$. Since the maximum $NPC_\alpha$ for each of the three models are below the respective $\overline{NPC}$'s, this means that I simply sampled the 100 eigenvectors with the largest $NPC_\alpha$ (the range of $NPC_\alpha$ sampled here are 386-441, 512-701 and 294-431 for the 3D, 1D and 1D NN, respectively). An additional constraint that I enforced is that $NPC_i > 100$ so that we avoid rows that have atypical amount of structure since those are peculiarities of the particular basis we are working in as well as for the fact that unfolding rows with low $NPC_i$ may not be very reliable for obvious reasons.

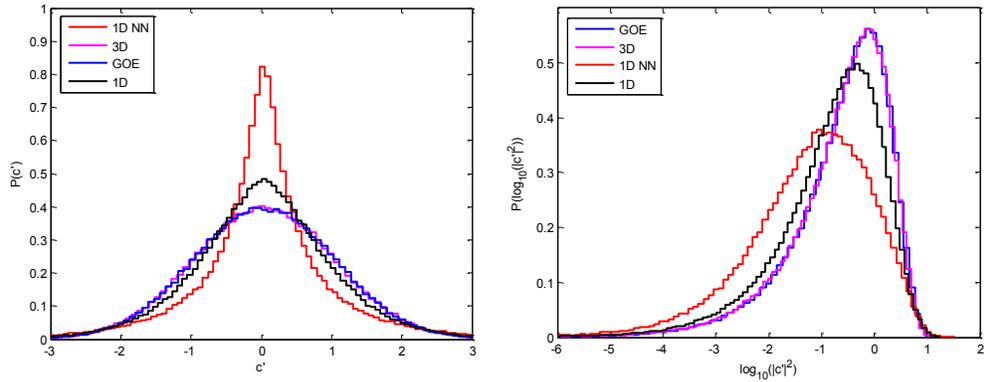

Figure 13: These plots were produced by adhering to the criteria discussed above (the number of components sampled 148800, 234000 and 182600 for the 3D, 1D N=15 and 1D NN N=15, respectively).

Figure 13 shows the eigenvector component statistics obtained by adhering to the criteria discussed above. The plot on the left is simply the normalized-component distribution. If the distribution is a normal distribution, as followed by the histogram for a GOE matrix (in blue) above, that implies that the eigenvector components are completely uncorrelated and random. On the other hand, the plot on the right is a histogram of the logarithm of the component-squares for the same components as the plot on the left; the distribution corresponding to full chaoticity is the Xi-square distribution with one degree of freedom, also known as Porter-Thomas distribution [2], given by

$$P_{PT}(x) = e^{-x/2}/\sqrt{x}$$

Note how the 3D histograms lines up almost exactly with the GOE ones, while the 1D histogram diverges somewhat but still clearly leans towards the GOE as opposed to the 1D NN. The closeness of the agreement of the 3D model with GOE prediction is remarkable and reconfirms the conclusion of our earlier analysis of eigenvalues that the 3D model is highly chaotic.

# Chapter 3
# Local Observables: Static Properties

The next topic in our research involves investigating the statistical properties of quantum mechanical observables – physical quantities that can, in principle, be measured experimentally. We would like to study, in particular, the fluctuation of expectation values of observables about their mean values, and examine the extent to which the fluctuations follow a normal distribution. But first we must identify other dependencies and rescale the data accordingly, similar to what we did for eigenvector components.

Our 1D model has a Hilbert space $\mathcal{H} = \otimes_{i=1}^{N} \mathcal{H}_i$, where the $\mathcal{H}_i$'s represent the Hilbert spaces of each spin-1/2 nuclear spin. In the thermodynamic limit ($N \to \infty$), the equipartition theorem would suggest that the total energy of a system is shared equally among each "degrees of freedom". Each spin is clearly a separate degree of freedom, thus the following simplified argument holds. Let $|E, N\rangle$ be an energy eigenstate corresponding to energy eigenvalue $E$ of the full system. Consider a subsystem $\mathcal{H}_m$ of $m$ spins, $m \leq N$; denote the Hamiltonian of the subsystem as $H(m)$. Then, the energy of the subsystem is given by (see [5] for further discussion)

$$\langle E, N | H(m) | E, N \rangle \approx \frac{m}{N} E =: \bar{E} \tag{3.1}$$

where $\bar{E}$ is the classical average energy of the subsystem. Furthermore, it can be shown that for any observable $A$ on $\mathcal{H}_m$,

$$\frac{1}{N_\epsilon} \sum_{\bar{E} < \frac{m}{N} E < \bar{E} + \epsilon} \langle E, N | A | E, N \rangle \approx A_{therm}(\tilde{E}) \tag{3.2}$$

where $N_\epsilon$ is the number of eigenstates in the energy window. In other words, the *average* of the expectation value of $A$ over a small enough energy window $(\bar{E}, \bar{E} + \epsilon)$ will converge to the classically-calculable value $A_{therm}(\tilde{E})$, also called the *microcanonical average* in statistical ensemble theory. Mathematically, equation (3.2) is defines *weak convergence*. Note that we made no assumption about the integrability of the system; both integrable and chaotic quantum systems are required to satisfy weak convergence [5].

We now present results on statistical properties of the 1D and the 3D models in relation to the expectation value of two types of local observables: the z-component of the i$^{th}$ spin $\sigma_i^z$, a one-body local observable, and the interaction energy $H_{ij}$ between a spin pair $(i, j)$, a two-body local observable

## 3.1 Local Spin Observable

Figure 14, the dots on the plot on the left show the expectation value $\langle \sigma_4 \rangle = \langle E_\alpha | \sigma_4 | E_\alpha \rangle$ of the fourth spin on the 1D chain as a function of energy $E_\alpha$, while the plot on the right shows the expectation value of one of the corner spins of the 3D lattice. As with eigenvector components, we would need to rescale the variance of expectation value of the local observable to adjust for its dependence on energy. In addition, however, $\langle \sigma_4 \rangle$ has a linearly-varying component (*i.e.* its *mean* varies along the energy spectrum). Applying our self-written kernel smoother, the linearly-varying component appears to be a horizontal line, *i.e.* a small (positive) constant. This is in accordance with what we anticipated since these results correspond to the one symmetry sector hence the microcanonical value for the spin is the same across the spectrum.

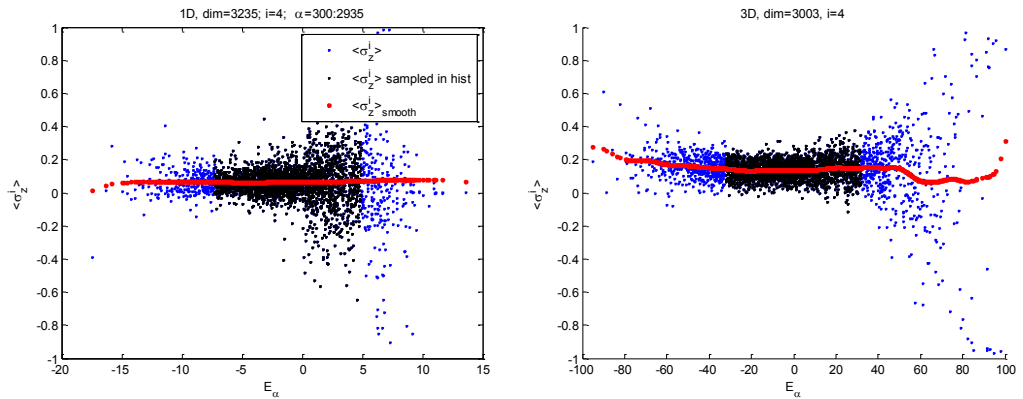

Figure 14: Expectation of local spin operator for the 4$^{th}$ spin of the 1D chain (left), and a corner spin for the 3D model (right). The horizontal line is around 1/15 for the former and 2/14 for the latter, corresponding to $S_z = +1$ and $S_z = +2$ blocks, respectively.

However, Figure 14 is curious in that we see an abrupt rise in the variance starting at $E = 0$ for the 1D model. This may be reminiscent of the asymmetry in the smoothness of the level density histogram at the beginning of chapter 3. Could there be some kind of phase transition at $E = 0$ for the 1D model? It is an exciting question that requires more study. As a check, I computed the corresponding plot for the 3rd and 5th as well as 8th (*i.e.* central) spins in the chain and we do see this abrupt rise in variance for those spins as well:

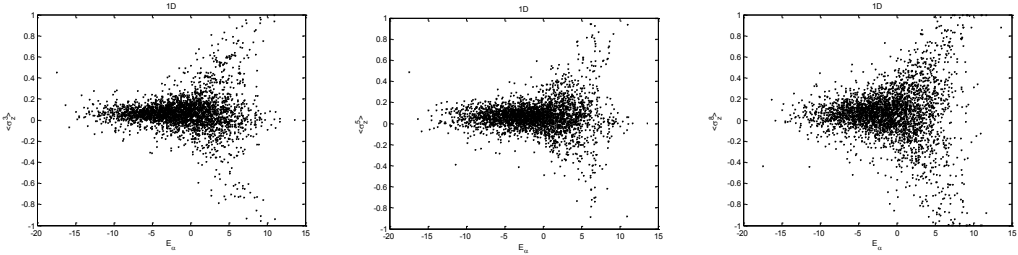

<u>Figure 15:</u> Expectation of local spin operator for the 3rd, 5th and 8th (*i.e.* central) spin of the N=15 1D chain. The sharp rise in variance around E=0 seem to exist for all spins.

We plan to come back to this result in future research. For the time being, we proceed with generating the distribution obeyed by the expectation value; we subtract off the linearly-varying component:

$$\langle \sigma_i^z \rangle' = \langle \sigma_i^z \rangle - g(E_\alpha)$$

and plot the square of $\langle \sigma_i^z \rangle'$ as shown in Figure 16.

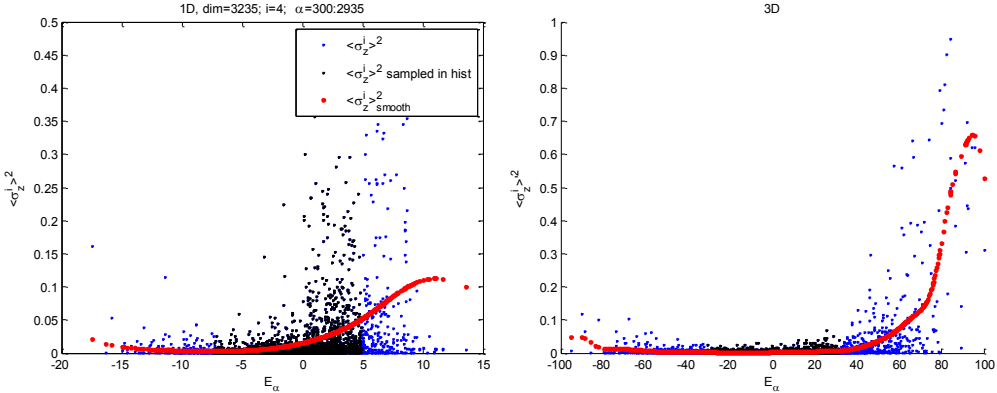

<u>Figure 16:</u> Plot of modulus-squared of $\langle \sigma_i \rangle'$ against energy for the 1D (left) and 3D (right) models.

Figure 16 shows how the variance of the (translated) expectation value varies with energy. We see the rescale this data to have unit variance

$$\widetilde{\langle \sigma_i^z \rangle}' = \frac{\langle \sigma_i^z \rangle'}{\sqrt{w(E^\alpha)}}$$

Note that the blue dots in the above figures represent regions where the kernel smoother appears to be unreliable. Thus, we sample only the black dots, generating the histograms shown in Figure 17.

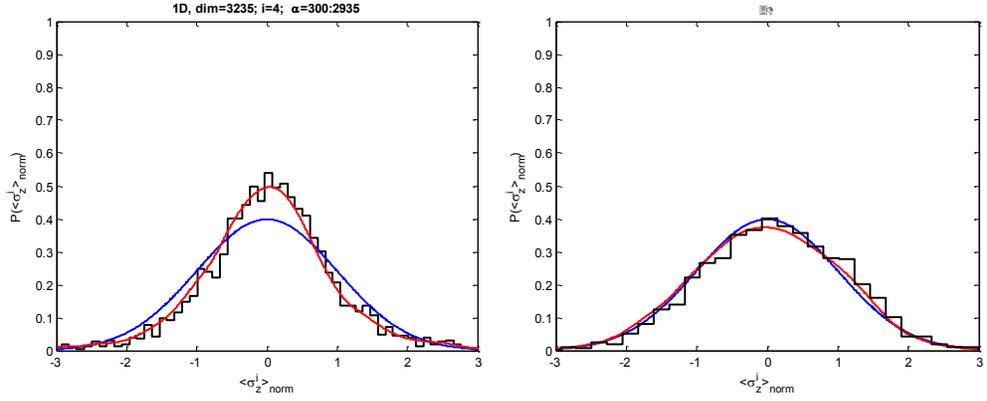

Figure 17: Histogram of $\widetilde{\langle \sigma_i^z \rangle}'$ for the 1D (left) and 3D (right) models. The curves in red are kernel-smoothed densities given by Matlab routine *ksdensity*.

The distribution of expectation value of local spin for the 1D model shown in Figure 17 appears to be somewhat closer in shape to a normal distribution than what we saw for eigenvector components in the previous chapter. On the other hand, the 3D model once again demonstrates that it is clearly chaotic.

## 3.2  Two-Body Observable

The observable $H_{ij}$ denotes the energy contribution from the interaction between the spin pair $(i,j)$. The dots in Figure 18 below shows $\langle H_{ij} \rangle = \langle E_\alpha | H_{ij} | E_\alpha \rangle$ versus energy $E_\alpha$. Applying our custom-written kernel smoother, a smoothly-varying linear trend $g(E_\alpha)$ emerges (shown in red), implying that on average, $\langle H_{ij} \rangle$ is proportional to $E_\alpha$.

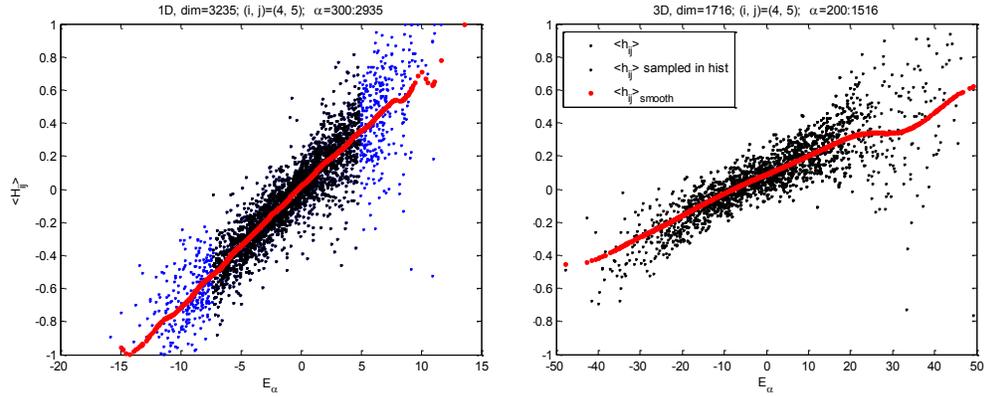

Figure 18: Expectation value of two-body energy contribution $\langle H_{ij} \rangle$ versus energy for the 1D (left) and 3D (right) models. These results correspond to $(i,j) = (4, 5)$ for the 1D model, which corresponds to the 4th and the 5th spins on the $N = 15$ chain, and two spins on adjacent corners of the cubic lattice for the 3D model. The blue dots represent regions is where the smoother appears to become somewhat inaccurate and are thus not sampled in the histograms below.

The linear trend is consistent with equation (4.1) above. We subtract off this linearly-varying component, viz.

$$\langle H_{ij} \rangle' = \langle H_{ij} \rangle - g(E_\alpha)$$

Next, note that we further since $H_{ij}$ is a quadratic degree of freedom (since it is a linear combination of products of two local spins), we expect from the equipartition theorem that the variance of $\langle H_{ij} \rangle'$ is a quadratic function of energy, $w(E_\alpha)$. Applying our smoothing routine on $\langle H_{ij} \rangle'^2$, we do see a quadratic-looking trend emerging as shown in the figure below.

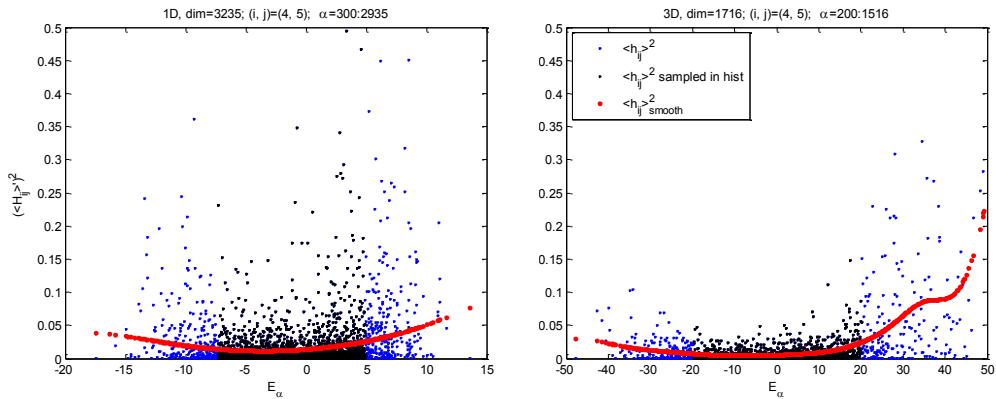

Figure 19: Plot of modulus-squared of $\langle H_{ij} \rangle'$ against energy for 1D (left) and 3D (right) models.

We rescale $\langle H_{ij}\rangle'$ as follows

$$\widetilde{\langle H_{ij}\rangle}' = \frac{\langle H_{ij}\rangle'}{\sqrt{w(E^\alpha)}}$$

Since the smoothed functions $f(E_\alpha)$ and $g(E_\alpha)$ appear to be less reliable near the edges, we chose to sample only the black dots in the central region.

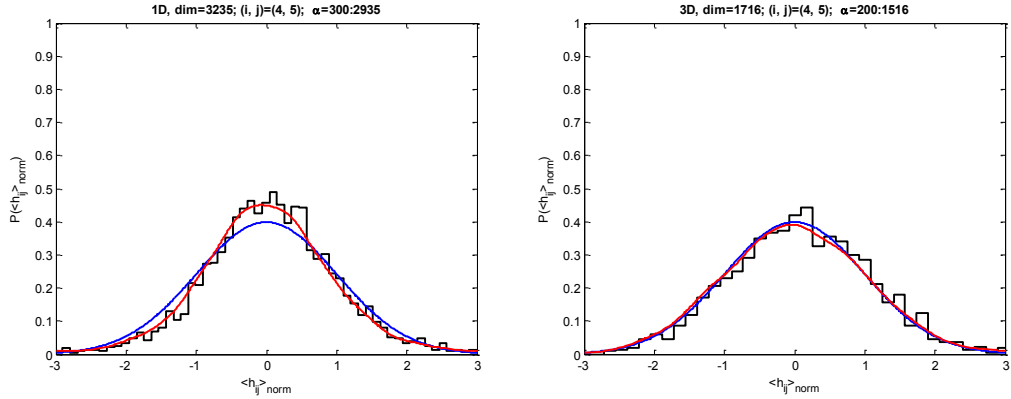

<u>Figure 20:</u> Histogram of $\widetilde{\langle H_{ij}\rangle}'$ for the 1D (left) and 3D (right) models. The curves in red are kernel-smoothed densities given by Matlab routine *ksdensity*.

The resulting distribution for the 1D model shown in **Figure 20** appears to be somewhat closer in shape to a normal distribution than what we saw in eigenvector statistics. On the other hand, the 3D model once again demonstrates it is chaotic.

## 3.3   Concentration of Measure

We showed in Section 3.1 that the expectation value of any local observable $A$ of a quantum mechanical system converges weakly to its microcanonical value in the thermodynamic limit. However, there is a stronger result for chaotic systems that was proposed in [5]. It is argues there that for chaotic Hamiltonians, the variance of $A$ within a small enough window must vanish in the thermodynamic limit. This regime is dubbed *strong convergence with universal fluctuations*.

A sketch of the argument in [5] is as follows. The essence of the application of RMT to chaotic Hamiltonians is that the components of chaotic energy eigenvectors are *as random as possible* once adjusted for other constraints. If the dimension of the Hilbert space is $d$, then the eigenvectors (which, in general, are complex numbers) are randomly chosen from the $(2d-1)$-dimensional hyper-sphere $S^{2d-1}$, analogous to the Bloch sphere for a single spin. The expectation value of any local observable is a real-valued function on this hyper-sphere; the probability that such a function takes a value larger than its mean value falls off exponentially with the dimension of the hyper-sphere [5]. This phenomenon is known mathematically as *concentration of measure,* which can be thought to generalize the well-known central limit theorem. Consequently, it can be shown that the variance of expectation values of a physical observable at a given energy scales as the reciprocal of the level density at that energy.

The plots in Figure 21 show the expectation value as a function of energy of a two-body observable $H_{45}$ representing the energy contribution from the 4th and 5th spins in the chain for spin chains of size $N = 13, 15, 17$ and $19$ respectively. Compare this set of plots for their counterparts in the nearest-neighbor limit shown in Figure 22. Note how in **Figure 21** the points in the plots concentrate around the microcanonical value for larger $N$. All four of the plots in **Figure 21** can be superimposed to further demonstrate this concentration pictorially, as presented in **Figure 23**. Note that such concentration of measure also happens for in plots for the integrable models but in a much less pronounced way.

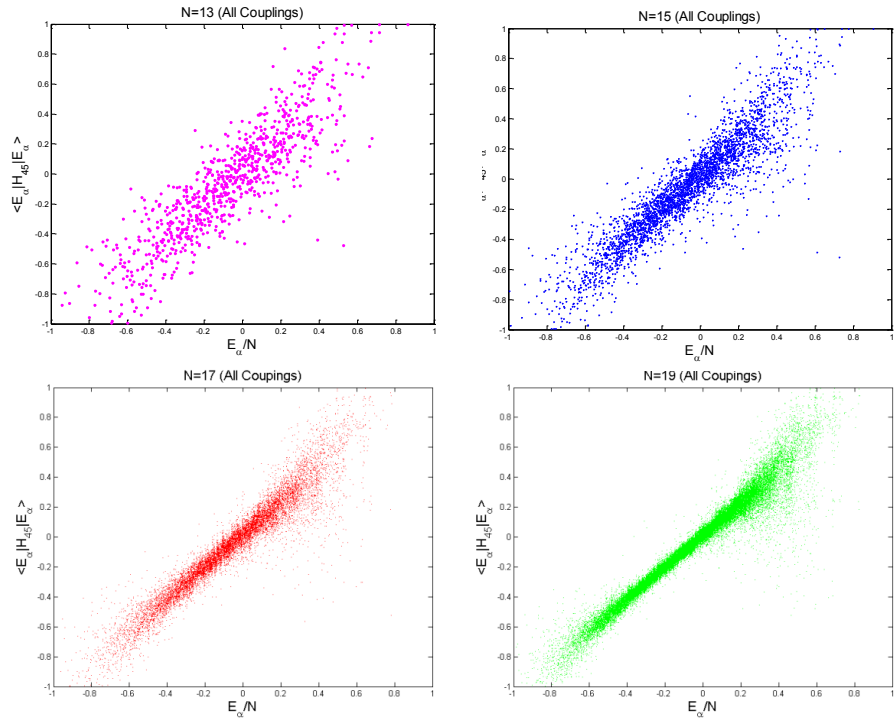

Figure 21: Expectation value of a two-body observable $H_{45}$ as a function of energy for the 1D model with spins N=13, 15, 17 and 19 respectively

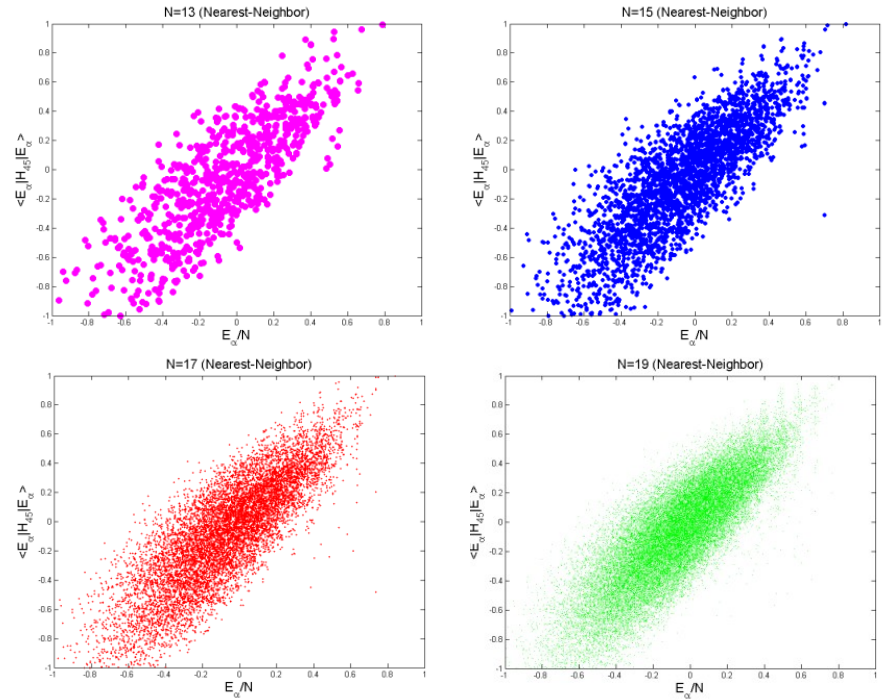

Figure 22: Expectation value of a two-body observable $H_{45}$ as a function of energy for the 1D NN model with spins N=13, 15, 17 and 19 respectively

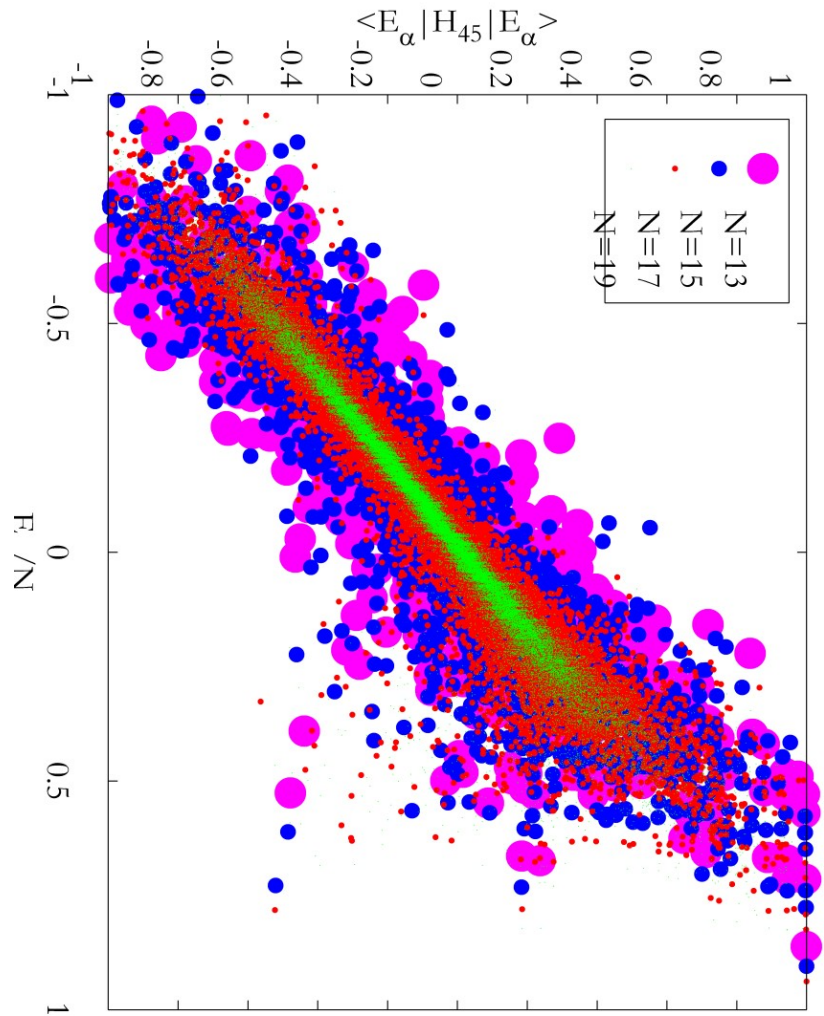

Figure 23: The plots in Figure 21 superimposed to visually illustrate the phenomenon of concentration of measure.

We quantified the rate of concentration for the 1D (chaotic) and 1D NN (integrable) models by plotting the natural log of the variance of $\langle H_{45} \rangle$ over a small energy window at the center of the spectrum against the number of spins $N$. On the same axes, we also plotted the natural log of the reciprocal of the level density at the center of the spectrum $E = 0$. As shown in **Figure 24** below, the two plots (the red and the black lines) are close to being parallel to each other, indicating that the variance $v$ of $\langle H_{45} \rangle$ scales (approximately) inversely with level density. That is, $v \sim 2^N$.

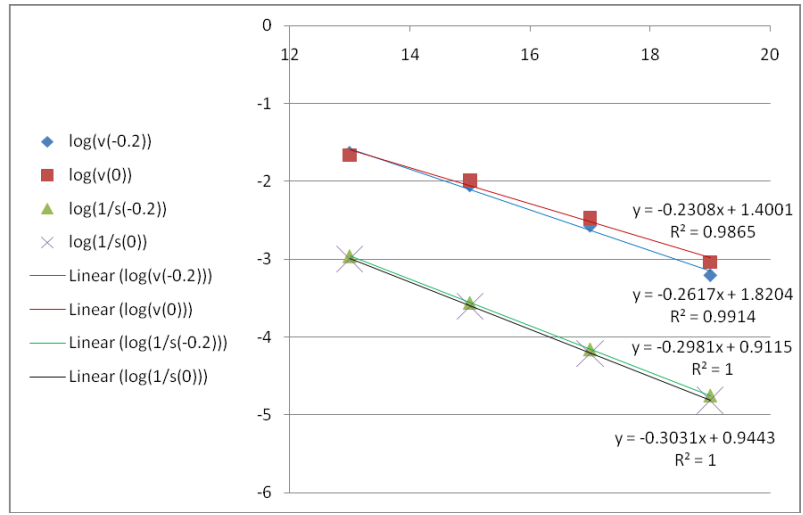

Figure 24: Concentration of measure analysis for the 1D (chaotic) model. The x-axis is number of spins $N$ and the y-axis is in the natural logarithm scale.

**Figure 24** is robust evidence of the high degree of chaoticity in the 1D model. We repeated this analysis at a point somewhat off-center of the spectrum, at $\frac{E}{N} = -0.2$, and found that the slopes of the variance and inverse density lines are even closer (blue and green lines, respectively), confirming (as we also saw with LSI as a function of energy in section 3.4) that the chaoticity is highest somewhere in the negative energy region rather than exactly at the center of the spectrum.

For comparison, we repeated analysis for the 1D NN model as well, shown in **Figure 25**. Note how much smaller the absolute value of the slope of $v$ is relative to that of the 1D model. The slope is still negative, meaning that concentration of measure is still occurring, but it is occurring at a *much slower rate* than for the chaotic case.

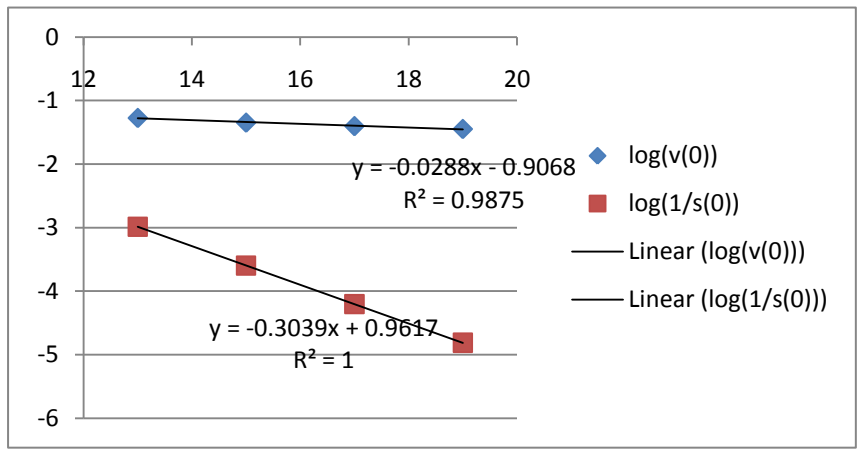

Figure 25: Concentration of measure analysis for the 1D NN (integrable) model. The x-axis is number of spins $N$ and the y-axis is in the natural logarithm scale.

# Chapter 4

# Time-Dependent Properties: Preliminary Results

The final topic in this thesis is a first look into time evolution of local observables. The observable we decided to focus on for this thesis is the local spin $\sigma_i^z$, where $1 \leq i \leq N$. Since we are already in possession of energy eigenvalues and energy eigenvectors from our static studies in the previous chapters, it is in principle straightforward to time-evolve any given initial state.

Whereas studying spectral properties required us to restrict ourselves to a single symmetry sector of a particular total spin $S_z$ subspace in order to avoid washing out the Wigner-Dyson distribution, the restriction to stay within a symmetry sector is no longer necessary while studying dynamics. We could, in principle, diagonalize all of the $S_z$ blocks of the full block diagonal Hamiltonian $H_D$ separately, and then time-evolve *any* computational basis state or linear combinations thereof. Indeed, being able to diagonalize one of the central blocks with available numerical resources implies that diagonalizing the full block-diagonal $H_D$ is also well-within our reach. However, since this is a preliminary study of dynamics, we will only time-evolve initial states within one $S_z$ subspace.

Suppose our chosen $S_z$ block is in some basis $\mathfrak{B}^{S_z} = \{|b_i\rangle\}_{i=1}^{\dim(S_z)}$, and let $|\psi_0\rangle$ be an initial state within the $S_z$ subspace. Working in the Schrödinger picture, $|\psi_0\rangle$ can be time-evolved, *i.e.* propagated, to time $t$ by application of the unitary propagator $e^{-iHt/\hbar}$. As before, it suffices to set $\hbar = 1$. The time-evolved state is therefore given by

$$|\psi(t)\rangle = e^{-iHt}|\psi_0\rangle = e^{-iHt} \sum_{\alpha=1}^{\dim(S_z)} |E_\alpha\rangle\langle E_\alpha|\psi_0\rangle$$
$$= \sum_{\alpha=1}^{\dim(S_z)} \left(e^{-iHt}|E_\alpha\rangle\right)\langle E_\alpha|\psi_0\rangle = \sum_{\alpha=1}^{\dim(S_z)} \left(e^{-iE_\alpha t}|E_\alpha\rangle\right)\langle E_\alpha|\psi_0\rangle$$

where $|E_\alpha\rangle$ corresponds to the energy eigenvector associated with energy eigenvalue $E_\alpha$. Note that if $V$ is the *modal matrix*, i.e. matrix whose columns are the energy eigenvectors in the basis $\mathfrak{B}^{S_z}$, then $V_{i\alpha} \equiv \langle b_i|E_\alpha\rangle$ and therefore $\langle E_\alpha|\psi_0\rangle = \sum_{i=1}^{\dim(S_z)}\langle E_\alpha|b_i\rangle\langle b_i|\psi_0\rangle = \sum_{i=1}^{\dim(S_z)} V_{\alpha i}^*\langle b_i|\psi_0\rangle$. Note that $V_{\alpha i}^* = V_{\alpha i}$ since $H_D$ is real in $\mathfrak{B}^{S_z}$, which in turn is real since the expression in equation (2.1) is real in the computational basis. Thus, we have

$$|\psi(t)\rangle = \sum_{\alpha=1}^{\dim(S_z)} \left(e^{-iE_\alpha t}|E_\alpha\rangle\right) \sum_{i=1}^{\dim(S_z)} V_{\alpha i}\langle b_i|\psi_0\rangle$$

After some rearrangement, we may write

$$|\psi(t)\rangle = \sum_{\alpha=1}^{\dim(S_z)} e^{-iE_\alpha t} [V^T|\psi_0\rangle]_\alpha |E_\alpha\rangle \tag{4.1}$$

where $V^T|\psi_0\rangle$ is the matrix multiplication of the transpose of the modal matrix with the column vector $|\psi_0\rangle$ in basis $\mathfrak{B}^{S_z}$. Finally, the expectation value of local spin $\sigma_i^z$ at time $t$ is

$$\langle\sigma_i^z(t)\rangle := \langle\psi(t)|V^T\sigma_i^z V|\psi(t)\rangle \tag{4.2}$$

where the observable $\sigma_i^z$, which is initially in basis $\mathfrak{B}^{S_z}$, is transformed to be in the eigenbasis before the expectation value is calculated.

## 4.1 1D XX Model

For reasons that will soon become clear, I first present time-evolution results for the case when the Ising term in our expression for the 1D Dipolar model in equation (1.3) is set to zero. We will call this the **1D XX model**. Figure 26 below depicts time evolutions of expectation values of local spin observables in the $N=15$ 1D XX model. One of the curves (the lightest one) corresponds to the NN (integrable) limit while the green curve is its long-range counterpart; the rest of the curves also correspond to the long-range model but for different initial states or spin locations.

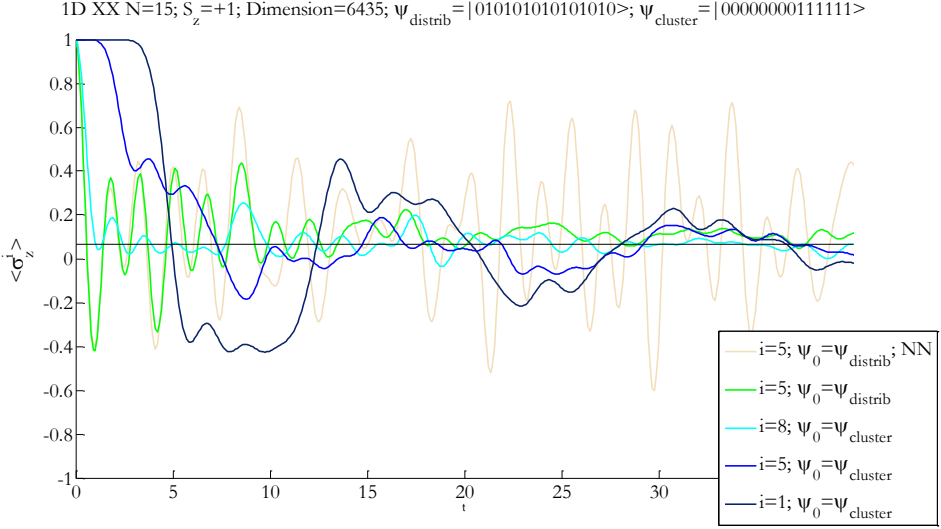

Figure 26: Local spin dynamics for the N=15 1D XX model.

Note how the green curve and its NN counterpart overlap initially and decay and oscillate together. This is a manifestation of the $\frac{1}{r^3}$ spatial dependence which causes the interaction strength between pairs of spins to drop rapidly with inter-spin distance. As a result, the nearest-neighbor interaction strength dominates over the longer-range couplings initially for a short period of time.

Eventually, the effects of the long-range couplings become very prominent. The black chaotic curve dissociates from the gray integrable curve and decays to a (nearly) steady state (with small fluctuations), in contrast with the persistent large oscillations of the integrable curve. Note that the small fluctuations for the chaotic case are a result of complicated quantum mechanical interactions that are not solvable analytically. However, the steady state that the chaotic case converges to in the long-time limit can be accurately predicted. All the long-range curves reach around $1/15$, which is exactly the microcanonical average for this subspace since there are 8 up-spins and 7 down-spins corresponding to an average spin of $1/15$ per spin in thermal equilibrium. This phenomenon of expectation value of local spins reaching a steady state is sometimes referred to as **equilibration**. If the steady state is at the predicted microcanonical value in the long-time limit, we refer to it as **thermalization**.

The different combinations of $i$ and the two disparate initial states $|010101010101010\rangle$ and $|000000001111111\rangle$ demonstrate that the behavior of the 1D XX is quite robust in the sense that all spins on the chain appear to thermalize for all initial states, albeit at somewhat different rates (at least for finite $N$ as considered). Shown in Figure 27 is support distribution of the two initial states over the eigenbasis, defined as

$$P(E) = \sum_{\alpha:\, E_\alpha \in [E, E+\epsilon]} |c_\alpha|^2$$

which has been smoothed using *ksdensity*.

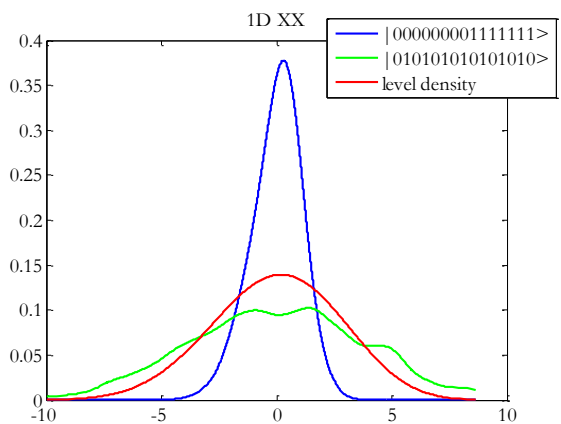
Figure 27: Support density of the clustered state (blue) and distributed state (green); the level density is given in red for comparison.

## 4.2 3D Dipolar Model

Having benchmarked our ideas about local spin dynamics and thermalization with the 1D XX model, we are well-prepared to attack the dipolar models, beginning with the 3D geometry. Firstly, note that in the 1D geometry, the one's and zeros' in the bit strings for each initial state are in their *actual position* on the chain; thus,

$|010101010101010\rangle$ and $|000000001111111\rangle$ are "special" since these two bit strings are the two opposite extremes of "clustering" of like-spins on the chain.

However, for the 3D lattice, there are few (if any) such strong clustering configurations possible. Thus, I took a different approach in choosing which initial state to time-evolve for the 3D model than for the 1D geometry. I chose as the initial state the basis element $|b_j\rangle \in \mathfrak{B}^{S_z}$ for which the following quantity is largest: $\xi(|b_i\rangle) = (\sum_\alpha |\langle E_\alpha | b_i \rangle|^4)^{-1}$. In other words, I chose the initial state whose projection onto the eigenbasis has the largest NPC, as defined earlier in the thesis.

Figure 28 below shows the time-evolution for the highest-NPC basis state, with $NPC = 812$ (dimension of subspace is 3003). It appears that all $N = 14$ spins thermalize to the microcanonical value 2/14 for the $S_z = +2$ block, although the "rogue spin" (shown in magenta) thermalizes at a somewhat slower rate. Also shown in Figure 28 is the distribution of support over the eigenbasis (in green) compared with the level density (in red).

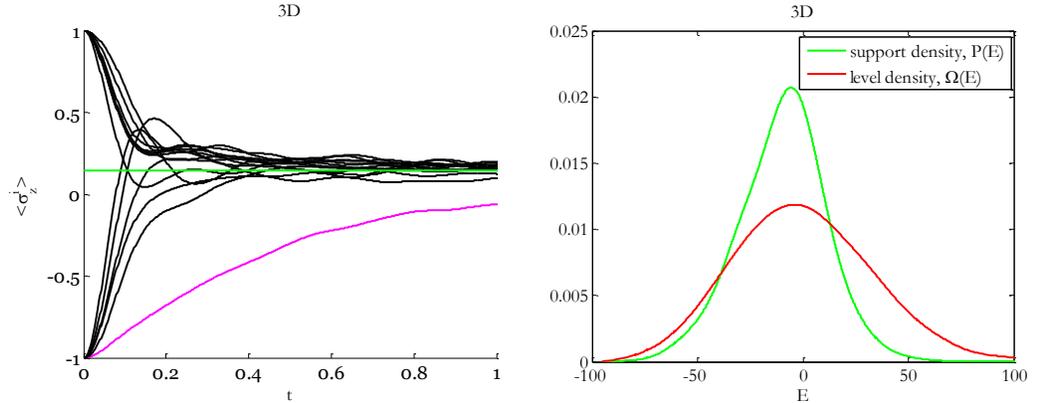

Figure 28: **(Left)** Local spin expectation values for all $N = 14$ spins for initial state $|b_j\rangle$. The magenta curve corresponds to the rogue qubit; the microcanonical value is in green. **(Right)** "Support Density" (computed using *ksdensity*) for the initial state $|b_j\rangle$ with highest NPC.

In summary, it appears that the 3D Dipolar model thermalizes at least for high NPC initial states. In future work, I plan to generate time-evolution plots for lower NPC initial states to investigate how robust the 3D Dipolar model is with respect to thermalization.

## 4.3 1D Dipolar

Finally, we take are ready to tackle the original question that was one of the key motivations for this research: the dynamical properties of the 1D Dipolar model. The first thing I tried was to run the same two initial states $|010101010101010\rangle$ and $|000000001111111\rangle$ that I tried for the 1D XX case, which we will call "distributed" and "clustered," respectively. The results for the 5th spin on the chain is shown in **Figure 29** below. Perhaps the most conspicuous and remarkable feature of the plot is that the clustered state (shown in blue) appears to stay completely stable at its initial value! We did not see such behavior in either the 1D XX or the 3D Dipolar models. The support density plot for the clustered state at the right of the figure (also shown in blue) reveals that the clustered state has almost all of its support at the high-energy end of the spectrum.

The evolution for the distributed state is somewhat less baffling since it follows the usual pattern of an initial decay followed by fluctuations about some average value. But it is still strange in that the fluctuations are somewhat large. Moreover, the fluctuations are similar in amplitude (perhaps even slightly larger) to the nearest-neighbor case (shown in lighter green)! We are not sure if the fluctuations are small enough to deserve to be called equilibration. The support density plot for the distributed state (shown in green) reveals that the distributed state has support mostly over the low-energy region of the spectrum.

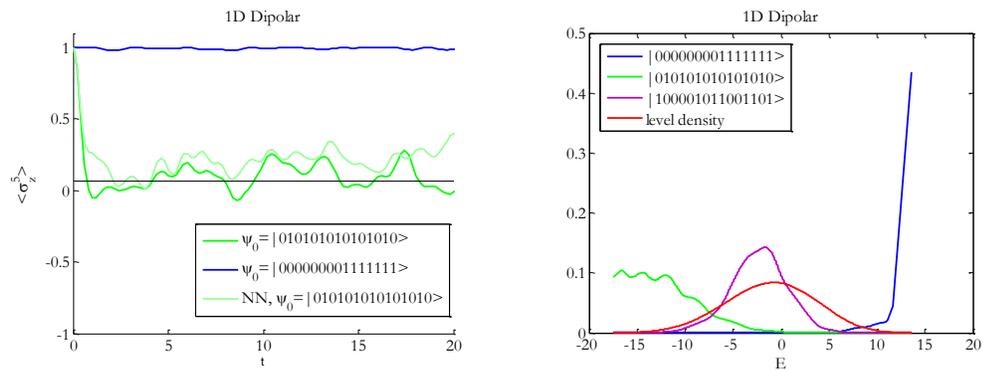

Figure 29: **(Left)** Time evolution results for local spin $\sigma_z^5$ in the N=15 1D Dipolar model. **(Right)** Support density of the clustered state (blue) and distributed state (green), as well as the highest NPC state $|100001011001101\rangle$ (magenta); the level density is given in red for comparison.

At first we were bewildered by the behavior of these two states, especially in light of their corresponding behavior of these exact same states and spin location for the 1D XX as shown in **Figure 26**. Indeed, I emphasize that the only difference while generating the plots for the 1D XX model and the 1D Dipolar model is that for the XX case I set the Ising term $\sigma_i^z \sigma_j^z$ in equation (1.3) to zero. Everything else is identical, yet the behavior of the two models is very different. But it may be helpful to note that perhaps these clustered and distributed states are somewhat atypical states for the 1D XXZ model as illustrated by their support densities peaking at the extremes of the energy eigenbasis. Why these states happen to be so pathological for the dipolar interaction is something we don't understand yet and a question that maybe deserve deeper investigation.

For the time being, however, let us look at a less atypical initial state, namely the state whose projection onto the eigenbasis has the highest NPC. For the 1D dipolar model, this state was computed to be the state |100001011001101⟩, with NPC=1127 (the dimension of the subspace is 6435). Its support density is shown in magenta in Figure 29. The time evolution of spins 1 through 8 (where spin-8 is, of course, at the center of the chain) for this initial state is given in Figure 30 below.

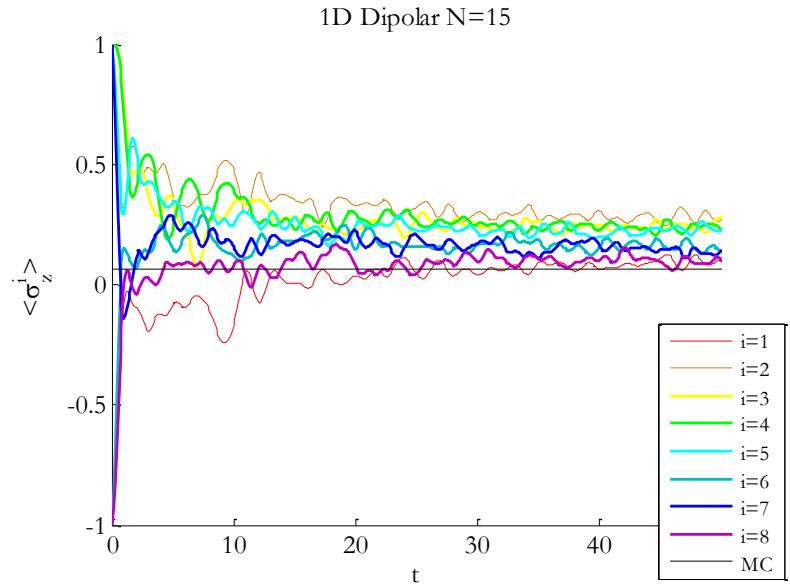

Figure 30: Time-evolution of the first 8 spins on the N=15 chain for the 1D dipolar model, with initial state |**100001011001101**⟩ that has the highest NPC=1127(the dimension of the subspace is 6435), corresponding to the magenta support density in Figure 27. The microcanonical value is shown in black.

The results in Figure 30 appear to be more in tune with our expectations than those in Figure 29. After the initial decay, all of the spins seem to settle down to some steady state with small fluctuations, *i.e.* we can say that all spins equilibrate. However, there appears to be a significant spread in the equilibration values of the different spins. A thinner line-width was used for Spin #1 and #2 to help distinguish them for the rest; perhaps the fact that these spins are the furthest away from the center of the spread is can be explained by boundary effects. A natural question to ask is whether this spread will get smaller as the size of the chain is increased. If it does, then that could be suggestive of thermalization for large enough chains.

We therefore close this section, and the chapter, with the time-evolution of expectation values of spins #1 through #10 of 1D Dipolar N=19 chain (*i.e.* spin #10 is the spin at the center of the chain) for its highest NPC basis state, $|1000010110011101\rangle$. Note, however, that this time we are only picking a basis from one symmetry sector of the $S_z = +1$ block since that is the largest matrix that can be diagonalized by our available computational resources. The results for $N = 19$ is shown in Figure 31 below.

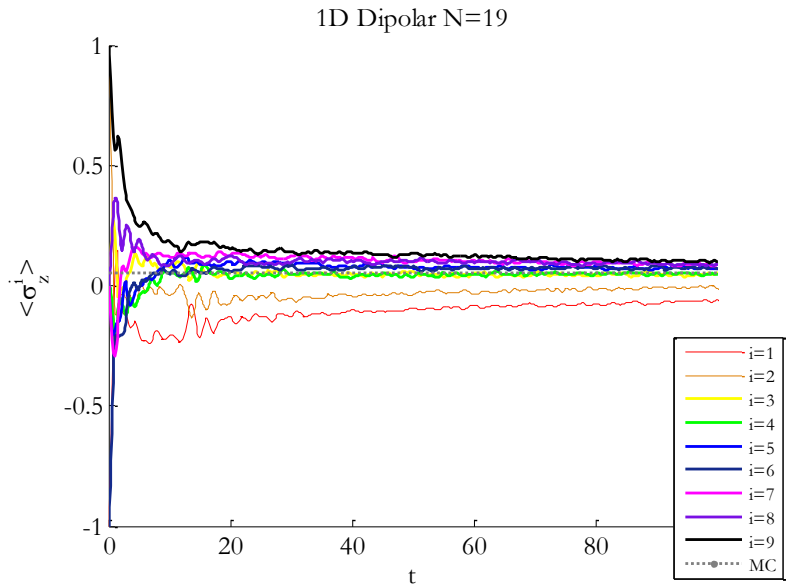

Figure 31: Dynamics of the first 10 spins on the N=19 chain for the 1D dipolar model, with initial state $\frac{1}{\sqrt{2}}(|1001110100001100101\rangle + |1010011000010111001\rangle)$ that has the highest NPC=1127 (the dimension of the subspace is 46252), corresponding to the magenta support density in Figure 32. The microcanonical value is shown in light gray.

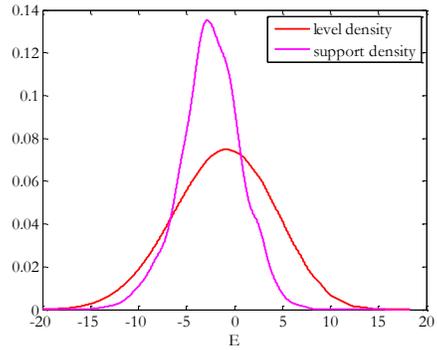

Figure 32: Support Density for initial state evolved in Figure 32.

As with the $N = 15$, we see a spread in the equilibration values of the 10 spins. However, we note that the spread for $N = 19$ is appreciably smaller, especially if we disregard the first two spins (shown in red and orange) as being irrelevant due to boundary effects. These are therefore encouraging signs that the 1D Dipolar chain does thermalize for large enough chains. However, to solidify this conclusion, we would need to run more simulations with other initial states – something that we plan to do in future work.

# Conclusion

We explored the statistical properties of eigenvalues, eigenvectors, local observables and time evolution of $N$ dipolarly-interacting homonuclear spins in (i) open chain, and (ii) face-centered lattice configurations. We began with a careful examination of eigenvalue statistics, obtaining reliable spacing distributions and LSI values via a proper unfolding procedure applied to the full spectrum. Next, we delved deeper into this class of chaotic spin-systems and studied some of its more detailed features, such as statistical features of energy eigenvectors and one- and two-body local observables. Finally, we presented some preliminary results on time-evolutions of local-spin observables.

In professional literature on the subject, eigenvector statistics has not been studied as extensively as eigenvalue statistics. Hence, my research involved not only the analysis of eigenvector statistics of the three models, comparison with GOE predictions and drawing conclusions about their chaoticity, but also exploring and benchmarking reliable numerical techniques to achieve these results. In particular, we demonstrated how to meaningfully extend the idea of unfolding, *i.e.* removing smoothly-varying dependencies, to eigenvector components and expectation value of observables.

Our analyses of eigenvalues, eigenvectors and local observables appear to be overall consistent with each other: We can collectively conclude that the 3D model appears to be fully chaotic whereas the 1D model is a somewhat intermediate case for smaller chains but approaches full chaoticity (*i.e.* LSI~0) for larger chains. One notable feature of the 1D Dipolar model is the asymmetry about the center of the energy spectrum that we observed in various quantities, such as the level density, NPC and the expectation value of local spins. Finally, our results on time-dependent properties seem to indicate that the isotropic 1D XX model thermalizes robustly, that is, for all locations on the chain and all initial states. The 3D Dipolar model appears to thermalize for most initial states. Finally, our preliminary look at the 1D Dipolar model indicates that it seem to thermalize for typical initial conditions.

While the problem of "many-body quantum chaos," and its possible link with quantum thermalization, is a fundamental open problem in quantum theory, our research is also directly relevant to quantum information science (QIS), in particular from the standpoint of building a quantum simulator on solid-state spin devices where dipolar couplings are naturally found in Nature.

# Appendix: Numerical Methods and Code

All of the coding for this project was done in Matlab (version 2009b) on my personal computer Windows XP (Dell Latitude D630). Nested calls to Matlab's *kron* routine were used to evaluate the tensor product between raising, lowering and/or identity operators and thereby form the full Hamiltonian matrix. Matlab's sparse matrix data structure made it possible to overcome the memory and processing challenges involved in handling square matrices of length of $2^{15} = 32768$ and larger. Overall, my code appears to be highly efficient, taking about 2 seconds for N=14 and 7 seconds for N=15 to generate the symmetry sector matrix.

Note that the size of the full Hilbert space for $N$ spins-1/2 spins grow as $2^N$, and the time required to fully diagonalize a Hamiltonian matrix of size M scales approximately as $M^3$. Consequently, memory requirements scale as $2^N$ and diagonalization time scales as $8^N$. Thus, Matlab's *eig* routine, which is responsible for diagonalizing the symmetry sector generated by my code, creates a bottleneck and imposes a limit on how large an N can be studied. *eig* takes around 20 seconds to diagonalize the subspace of dimension 1716 for the 3D lattice with N=14, and more than 3 minutes to diagonalize the subspace of dimension 3235 for the 1D chain with N=15.

These computational challenges put a constraint on the largest N that we can study on a personal computer. The highest N supportable on my machine, N=15, had served our purposes well initially and revealed a number of interesting features. But eventually we branched out to computing resources on the Dartmouth campus. We utilized Research Computing's *Andes* cluster and have succeeded in pushing up to N=19, the maximum number of spins allowable by memory constraints on Andes. We diagonalized the symmetry sector of dimension $46250$; each copy of this matrix required $\sim 17$ GB of memory, and the diagonalization routine ran remotely on Andes for approximately 76 hours. Enclosed in the next pages is the collection of code that computes the Hamiltonian and generates the subspaces studied in this thesis.

```matlab
%Author: Dhrubo Jyoti '11
%djyoti11@gmail.com

%fluorapatite
function H=fluorapatite(N, nearest) %specify if nearest neighbor
%% SET LATTICE STRUCTURE
x=0:1:N-1; y=zeros(1,N);
%figure; title('1D'); axis equal; hold on; scatter(x, y,'filled');
cij=sparse(N,N);

%% PARAMETERS
p=3; %long-range dependence
thetaij=pi/2; cosine=0; %1D geometry

%% FIND COEFFICIENTS
for i=1:N
    for j=i+1:N
        if  nearest==0 || j-i==1
            rij=x(j)-x(i);
            cij(i,j)=(3*(cosine^2)-1)/(2*rij^p); %NOTE: Constant factor gamma set to 1.
        end
    end
end

%% FIND CENTRAL BLOCK OF TOTAL HAMILTONIAN
H=evaluateH(N,cij,1);
['subspace dimension: ',num2str(length(H))] %print size of block
['Real & Hermitian: ',num2str(isreal(H) && isequal(H,H'))] %make sure its real and Hermitian
end
```

```matlab
%Author: Dhrubo Jyoti '11
%djyoti11@gmail.com

%Finds full HD matrix, extracts central block and returns
%desymmetrized sub-block.

function H=calciumfluoride(N)
%% SET LATTICE STRUCTURE AND PLOT IT. NOTE: normalized units
%CaF2 with one displaced qubit
x=[-0.15    0    0    0  0  1   1    1    1 .5 .5 .5 .5];
y=[-0.3     0   .5    1  1  0   0   .5    1 .5 .5  1  0];
z=[0        1   .5    0  1  0   1   .5    0  1  0  1 .5 .5];

figure('Position',[100 100 400 300]); axis equal; xlabel('x'); ylabel↵
('y'); zlabel('z');
title('Face-centered cubic lattice');
fill3(xx,yy,zz,'c','FaceColor','b','FaceAlpha',0.25); hold on;
scatter3(x(2:N), y(2:N), z(2:N),150,'filled','g'); hold on;
scatter3(x(1), y(1), z(1),150,'filled','r');
annotation('textarrow',[-0.15 -3 0]);

cij=sparse(N,N);

%% FIND COEFFICIENTS
tic;
HD=sparse(2^N,2^N);
for i=1:1:N
for j=1:1:N
    if i<j
        rij=sqrt(((x(i)-x(j))^2)+((y(i)-y(j))^2)+((z(i)-z(j))^2));
        cosine=(z(i)-z(j))/rij;
        cij(i,j)=(3*(cosine^2)-1)/(rij^3); %NOTE: Constant factor gamma↵
set to 1.
    end
end
end
toc;

%% FIND CENTRAL BLOCK OF TOTAL HAMILTONIAN
H=evaluateH(N,cij,0);
['subspace dimension: ',num2str(length(H))] %print size of block
%sanity check: make sure Hamiltonian is real and Hermitian:
```

```
['Real & Hermitian: ',num2str(isreal(H) && isequal(H,H'))]
end
```

```matlab
%Author: Dhrubo Jyoti '11
%djyoti11@gmail.com

%Finds dipolar Hamiltonian (H_D) matrix, extracts central block and↲
returns
%desymmetrized sub-block.
function H=evaluateH(N,cij,isFluorapatite)
tic;
HD=sparse(2^N,2^N);
for i=1:N
    for j=i+1:N
        h=cij(i,j)*evaluatehij(i, j, N);
        HD=HD+h; clear h;
    end
end
toc;

H=obtainSubspace(N, HD, isFluorapatite);
%NO NEED TO CALCULATE H_Z SINCE B IS UNIFORM; see my DUJS paper for↲
details.
end
```

```matlab
%Author: Dhrubo Jyoti '11
%djyoti11@gmail.com

%Computes two-body Hamiltonian for spin pair (i,j). N.B. Does not include
%interaction strength coefficient cij
function hij=evaluatehij(i,j,N)
if (i>=j)
    error('i must be strictly less than j');
end

%% SET COEFFICIENTS
J_x=1;
J_y=1;
J_z=-2;

%% PAULI MATRICES
zspin=sparse([1 0;0 -1]);
xspin=sparse([0 1;1 0]);
yspin=sparse([0 complex(0,-1); complex(0,1) 0]);

%% COMPUTE h_ij
leading=i-1;
sandwitch=j-i-1;
trailing=N-j;

if leading==0; leading=1; else leading=speye(2^leading); end
if sandwitch==0; sandwitch=1; else sandwitch=speye(2^sandwitch); end
if trailing==0; trailing=1; else trailing=speye(2^trailing); end

XX=kron(kron(kron(kron(leading,xspin),sandwitch),xspin),trailing);
YY=kron(kron(kron(kron(leading,yspin),sandwitch),yspin),trailing);
ZZ=kron(kron(kron(kron(leading,zspin),sandwitch),zspin),trailing);

hij=J_x*XX+J_y*YY+J_z*ZZ;
end
```

```matlab
%Author: Dhrubo Jyoti '11
%djyoti11@gmail.com

%Takes an operator on full Hilbert space, extracts central block
%and returns desymmetrized sub-block.
function operator=obtainSubspace(N, operator, isFluorapatite)
%% (1) Operator is assumed to be in computational basis
%ispy(operator);

%% (2) BLOCK DIAGONALIZATION
%Permute basis elements to arrive at block-diagonal form
basis=orderedBasis(N);
U=sparse(2^N,2^N); %Compute change of basis matrix
for i=1:1:2^N;
    U(find(basis+1==i),i)=1;
end
operator=U*operator*U'; %Apply change of basis matrix
%ispy(operator);

%% (3) EXTRACT A CENTRAL BLOCK
%Initialize some variables
blockDim=0; %Dimension of subspace
start=0;
finish=0;

if  mod(N,2)==0 %Choose Sz=+2 block for even number of qubits
    blockDim=nchoosek(N,1+floor(N/2)); start=0;
    for i=0:floor(N/2)-2 start=start+nchoosek(N,i); end; start=start+1;
    finish=start+blockDim-1;
else
    %Choose Sz=+1 block for odd number of qubits
    blockDim=nchoosek(N,floor(N/2)); start=0;
    for i=0:floor(N/2)-1 start=start+nchoosek(N,i); end; start=start+1;
    finish=start+blockDim-1;
end

%Extract the central block
v=operator(:,start:1:finish);
v=v(start:1:finish,:);
operator=v;
blockBasis=uint32(basis(start:1:finish));
%ispy(operator);
```

```matlab
%% (4) DESYMMETRIZE W.R.T. MIRROR SYMMETRY (For FLUORAPATITE only)
if isFluorapatite
    tic
    %Reorder basis elements into two lists "symm" and "antisymm" such
    %that one list maps to the other under the spatial "mirror
reflection"
    %transformation, and vice-versa.
    %Some basis elements are invariant; group them under "symm".
    
    symm=[]; antisymm=[];
    complements=ireflect(blockBasis,N); %Mirror image of each basis
element.
    for i=1:blockDim
        if blockBasis(i)==complements(i);
            symm=cat(2,symm,blockBasis(i));
        elseif nnz(symm==blockBasis(i))==0 && nnz(antisymm==blockBasis
(i))==0
            symm=cat(2,symm,blockBasis(i));
            antisymm=cat(2,antisymm,complements(i));
        end
    end
    newBasis=cat(2, symm, antisymm);
    U=sparse(blockDim,blockDim);
    for i=1:blockDim; U(find(newBasis==blockBasis(i)),i)=1; end
    operator=U*operator*U';
    
    blockBasis=newBasis;
    complements=ireflect(blockBasis,N); %size(blockBasis)
    U=sparse(blockDim,blockDim); NORM=1/sqrt(2);
    for i=1:blockDim;
        j=find(complements(i)==blockBasis);
        if   i<j;
            U(i,i)=NORM; U(i,j)=NORM; U(j,i)=NORM; U(j,j)=-NORM; %set
appropriate values
        elseif i==j;
            U(i,i)=1;
        end
    end
    
    operator=U*operator*U'; %apply basis transform; get two sub-blocks.
```

```matlab
    %ispy2(operator);
    symmdim=length(symm);
    operator=operator(:,1:symmdim); %pick the symmetric sub-block (upper
one) and return it
    operator=operator(1:symmdim,:);
    %ispy(operator);
    toc
end

end

%HELPER FUNCTION Takes in integers, reflects binary expansions about
%center, and returns corresponding integers.
function result=ireflect(rowVecPosInts,bitstrlength)
b=dec2bin(rowVecPosInts,bitstrlength);
b=b(:,bitstrlength:-1:1);
result=bin2dec(b)';
end

%HELPER FUNCTION: Makes nice sparsity plots
function ispy2(operator)
l=length(operator);
for i=1:l;
    for j=1:l; %remove round-off error
        if abs(operator(i,j))<10^-14 && abs(operator(i,j))>0; operator(i,
j)=0; end;
    end;
end;
operator=sparse(full(operator));
s=figure('Position',[100 100 340 300]);
set(s, 'DefaultTextFontSize', 10); set(s, 'DefaultTextFontName',
'Garamond');
set(s, 'DefaultAxesFontSize', 10); set(s, 'DefaultAxesFontName',
'Garamond');
spy(operator,0.5,'k');
set(gca,'xtick',[],'ytick',[]);
xlabel(['dim=',num2str(length(operator))]);
ylabel([]); title([]);
save2word('operator3.doc');
end

%HELPER FUNCTION: Makes nice sparsity plots
```

```matlab
function ispy(operator)
s=figure('Position',[100 100 340 300]);
set(s, 'DefaultTextFontSize', 10); set(s, 'DefaultTextFontName',↵
'Garamond');
set(s, 'DefaultAxesFontSize', 10); set(s, 'DefaultAxesFontName',↵
'Garamond');
spy(operator,0.5,'k');
set(gca,'xtick',[],'ytick',[]);
xlabel(['dim=',num2str(length(operator))]);
ylabel([]); title([]);
end
```

```matlab
%Author: Dhrubo Jyoti '11
%djyoti11@gmail.com

%Reorders the standard basis in ascending order of on-bits, i.e.
%down-spins. In other words, takes a list (i.e. row vector) of (unsigned)
%integers and sorts them in ascending according to the number of ones
%(i.e. down-spins) in their binary representation.
% We expect the Hamiltonian to be block diagonal in this basis.
function basis = orderedBasis(N)
basis = uint32(0:1:2^N-1); onCounts=uint32(zeros(1,2^N));
binary=dec2bin(uint32(basis)')=='1';
for i=1:2^N
    onCounts(1,i) = nnz(binary(i,:));
end
list=[basis; onCounts]; list=list';
list=sortrows(list,2);
list=list';
basis=list(1,:);
end
```

```matlab
%Author: Dhrubo Jyoti '11
%djyoti11@gmail.com

% Returns the smoothed values of a function specified by its
% domain set "x" and noisy value set "f". Accepts "window" parameter,
% or attempts to determine good window size itself via ksdensity.
% Optionally plots the noisy points and smoothed values.
function f_s=ismooth(x, f, window, doPlot)
dim=length(x); f_s=zeros(1,dim);
if window==0 %if window size not provided...
    [a,b,window]=ksdensity(x,x,'weights',f); %...find "optimal" window
end
kernel = @(x,s) exp(-(x.^2)/(2*s(1,1)^2))/(s(1,1)*sqrt(2*pi)); %Gaussian kernel
for i=1:dim
    k=kernel(x-x(i),window); f_s(1,i)=dot(k,f)/sum(k); %apply kernel
end

if doPlot
    hold on;
    font='Garamond'; fontsize=14; s=gcf; set(gcf, 'DefaultTextFontSize',fontsize);
    set(gcf, 'DefaultAxesFontSize', fontsize);
    set(gcf, 'DefaultAxesFontName', font);
    set(gcf, 'DefaultTextFontName', font);
    plot(x,f,'k.','MarkerSize',1); hold on;
    plot(x,f_s,'r.','MarkerSize',6); xlabel(xlab); ylabel(ylab); title(titl);
else
    hold on; plot(x,f_s,'r.');
end
end
```

```matlab
%Author: Dhrubo Jyoti '11
%djyoti11@gmail.com

%Unfolds the energy spectrum specified by "eigenvalues"; optionally
%truncates a given number of eigenvalues at the beginning and end specified
%by "truncate". Optionally plots the spacing distribution.
function [spacings, nu]=unfoldvals(eigenvalues,truncate, doPlot) %eigenvalues must be in ascending order
P_G = @(x) exp(-x.^2/2)/sqrt(2*pi); %Normalized Gaussian
P_P = @(x) exp(-x); %Normalized Poisson
P_WD = @(x) (pi/2)*x.*exp(-(pi/4)*x.^2); %Normalized Wigner-Dyson
s0=0.4729; %Poisson-WD intersection

%Find smoothed level density distribution rho, and normalize it
[rhoy,rhox]=ksdensity(eigenvalues,eigenvalues);
rhoy=length(eigenvalues)*rhoy;
figure; hist(eigenvalues, 30); hold on; scatter(rhox,rhoy,'r.');
spacings=diff(eigenvalues);
for i=1:length(spacings)
    avg=2/(rhoy(i)+rhoy(i+1));
    spacings(i)=spacings(i)/avg;
end

if truncate~=0
    spacings=spacings(1+truncate:length(spacings)-truncate);
end

integral_P=nnz(spacings<=s0)/length(spacings)
integralP_P=quadgk(P_P,0,s0); %integrate Poisson distribution, 0 to s0
integralP_WD=quadgk(P_WD,0,s0); %integrate WD distribution, 0 to s0
nu=(integral_P-integralP_WD)/(integralP_P-integralP_WD) %LSI

%% PLOT
if doPlot
    figure;
    font='Garamond'; fontsize=14; s=gcf;
    set(s, 'DefaultTextFontSize', fontsize);
    set(s, 'DefaultAxesFontSize', fontsize);
    set(s, 'DefaultAxesFontName', font);
    set(s, 'DefaultTextFontName', font);
```

```matlab
        binsize=2*iqr(spacings)*(length(spacings)^(-1/3))
        bins=round((max(spacings)-min(spacings))/binsize)
        xxx=0:0.001:4; [yy,xx]=hist(spacings,bins); bin=(xx(2)-xx(1));
        yy=yy/(length(spacings)*bin); stairs(xx,yy,'Color','k', 'LineWidth',↵
2);
        xlim([0 4]); ylim([0 1]); hold on;
        plot(xxx,P_P(xxx),'LineWidth',2,'Color','b');
        hold on; plot(xxx,P_WD(xxx),'LineWidth',2,'Color','r');
        xlabel('$$\tilde{s}$$','Interpreter','latex');
        ylabel('$$P(\tilde{s})$$','interpreter','latex');
        set(gca,'xtick',[0 1 2 3 4],'ytick',[0 .2 .4 .6 .8 1]);
end
end
```

```matlab
%Author: Dhrubo Jyoti '11
%djyoti11@gmail.com

%Returns expectation value of an observable as a function of time
% t=0 to t=20. Follows SCHROEDINGER PICTURE.

%psi_0 is the initial state (a column vector in computational basis),
%Obs is the observable (in the computational basis) whose expectation
%values needs to be computed, E is the column vector of energy
eigenvalues,
%and V is the modal matrix (i.e. columns are energy eigenvectors i.e. the
eigenbasis)
function expectation_t=evolve(psi_0, Obs, E, V)
%% Constants
END_TIME=100;
STEPS=1000;
timedomain=linspace(0,END_TIME,STEPS); %Time span (t=0 to t=20)
i=sqrt(-1);

%% Change basis to eigenbasis
tic; Obs=Obs*V; toc;
tic; U=V'; toc;
clear V;
Obs=U*Obs;
psi_0=U*psi_0; %Project initial state from computational onto eigenbasis
clear U;
tic; psi_t=diag(psi_0)*exp(-i*kron(timedomain,E)); toc;

%% Find time-evolved state by applying time-evolution operator
tic; expectation_t=diag(psi_t'*Obs*psi_t); toc; %take inner product
clear Obs;
hold on; plot(timedomain,real(expectation_t),'k','LineWidth',0.5);
end
```